\documentclass[twocolumn]{aastex631}

\usepackage[T1, OT1]{fontenc}
\usepackage{hyperref}

\shorttitle{Low Stellar Obliquity in the Transiting Brown Dwarf System GPX-1}
\shortauthors{Giacalone et al.}

\accepted{for publication in the Astronomical Journal on September 9, 2024}
\graphicspath{{./}{figures/}}

\begin{document}

\title{The OATMEAL Survey. I. Low Stellar Obliquity in the Transiting Brown Dwarf System GPX-1}

\correspondingauthor{Steven Giacalone}
\email{giacalone@astro.caltech.edu}

\author[0000-0002-8965-3969]{Steven Giacalone}
\altaffiliation{NSF Astronomy and Astrophysics Postdoctoral Fellow}
\affiliation{Department of Astronomy, California Institute of Technology, Pasadena, CA 91125, USA}

\author[0000-0002-8958-0683]{Fei Dai}
\affiliation{Institute for Astronomy, University of Hawai`i, 2680 Woodlawn Drive, Honolulu, HI, 96822, USA}
\affiliation{Division of Geological and Planetary Sciences,
1200 E California Blvd, Pasadena, CA, 91125, USA}
\affiliation{Department of Astronomy, California Institute of Technology, Pasadena, CA 91125, USA}

\author[0000-0002-9849-5886]{J. J. Zanazzi}
\altaffiliation{51 Pegasi b Postdoctoral Fellow}
\affiliation{Department of Astronomy, University of California Berkeley, Berkeley, CA 94720, USA}

\author[0000-0001-8638-0320]{Andrew W. Howard}
\affiliation{Department of Astronomy, California Institute of Technology, Pasadena, CA 91125, USA}

\author[0000-0001-8189-0233]{Courtney D. Dressing}
\affiliation{Department of Astronomy, University of California Berkeley, Berkeley, CA 94720, USA}

\author[0000-0002-4265-047X]{Joshua N.\ Winn}
\affiliation{Department of Astrophysical Sciences, Princeton University, Princeton, NJ 08544, USA}

\author[0000-0003-3856-3143]{Ryan A. Rubenzahl}
\altaffiliation{NSF Graduate Research Fellow}
\affiliation{Department of Astronomy, California Institute of Technology, Pasadena, CA 91125, USA}

\author[0000-0001-6416-1274]{Theron W. Carmichael}
\altaffiliation{NSF Ascend Postdoctoral Fellow}
\affiliation{Institute for Astronomy, University of Hawai`i, 2680 Woodlawn Drive, Honolulu, HI, 96822, USA}

\author[0000-0002-0701-4005]{Noah Vowell}
\affiliation{Center for Astrophysics \textbar \ Harvard \& Smithsonian, 60 Garden St, Cambridge, MA 02138, USA}
\affiliation{Center for Data Intensive and Time Domain Astronomy, Department of Physics and Astronomy, Michigan State University, East Lansing, MI 48824,
USA}

\author[0000-0002-3239-5989]{Aurora Kesseli}
\affiliation{IPAC, Mail Code 100-22, Caltech, 1200 E. California Boulevard, Pasadena, CA 91125, USA}

\author[0000-0003-1312-9391]{Samuel Halverson}
\affiliation{Jet Propulsion Laboratory, 4800 Oak Grove Drive, Pasadena, CA 91109, USA}

\author[0000-0002-0531-1073]{Howard Isaacson}
\affiliation{501 Campbell Hall, University of California at Berkeley, Berkeley, CA 94720, USA}
\affiliation{Centre for Astrophysics, University of Southern Queensland, Toowoomba, QLD, Australia}

\author[0009-0008-9808-0411]{Max Brodheim}
\affiliation{W. M. Keck Observatory, Waimea, HI 96743, USA}

\author[0009-0000-3624-1330]{William Deich}
\affiliation{University of California Observatories, 1156 High Street, Santa Cruz, CA 95064}

\author[0000-0003-3504-5316]{Benjamin J.\ Fulton}
\affiliation{NASA Exoplanet Science Institute/Caltech-IPAC, California Institute of Technology, Pasadena, CA
91125, USA}

\author[0009-0004-4454-6053]{Steven R. Gibson}
\affiliation{Caltech Optical Observatories, Pasadena, CA, 91125, USA}

\author[0000-0002-7648-9119]{Grant M.\ Hill}
\affiliation{W.\ M.\ Keck Observatory, 65-1120 Mamalahoa Hwy, Kamuela, HI 96743, USA}

\author[0000-0002-6153-3076]{Bradford Holden}
\affiliation{University of California Observatories, 1156 High Street, Santa Cruz, CA 95064}

\author[0000-0002-5812-3236]{Aaron Householder}
\affiliation{Department of Earth, Atmospheric and Planetary Sciences, Massachusetts Institute of Technology, Cambridge, MA 02139, USA}
\affil{Kavli Institute for Astrophysics and Space Research, Massachusetts Institute of Technology, Cambridge, MA 02139, USA}

\author{Stephen Kaye}
\affiliation{Caltech Optical Observatories, California Institute of Technology, Pasadena, CA 91125, USA}

\author[0000-0003-2451-5482]{Russ R. Laher}
\affiliation{NASA Exoplanet Science Institute/Caltech-IPAC, 1200 E California Blvd, Pasadena, CA 91125, USA}

\author[0009-0004-0592-1850]{Kyle Lanclos}
\affiliation{W.\ M.\ Keck Observatory, 65-1120 Mamalahoa Hwy, Kamuela, HI 96743, USA}

\author[0009-0008-4293-0341]{Joel Payne}
\affiliation{W.\ M.\ Keck Observatory, 65-1120 Mamalahoa Hwy, Kamuela, HI 96743, USA}

\author[0000-0003-0967-2893]{Erik A. Petigura}
\affiliation{Department of Physics \& Astronomy, University of California Los Angeles, Los Angeles, CA 90095, USA}

\author[0000-0001-8127-5775]{Arpita Roy}
\affiliation{Astrophysics \& Space Institute, Schmidt Sciences, New York, NY 10011, USA}

\author[0000-0002-4046-987X]{Christian Schwab}
\affiliation{School of Mathematical and Physical Sciences, Macquarie University, Balaclava Road, North Ryde, NSW 2109, Australia}

\author[0000-0003-3133-6837]{Abby P. Shaum}
\affiliation{Department of Astronomy, California Institute of Technology, Pasadena, CA 91125, USA}

\author[0009-0007-8555-8060]{Martin M.\ Sirk}
\affiliation{Space Sciences Laboratory, University of California, Berkeley, CA 94720, USA}

\author{Chris Smith}
\affiliation{Space Sciences Laboratory, University of California, Berkeley, CA 94720, USA}

\author[0000-0001-7409-5688]{Guðmundur Stefánsson} 
\affil{Anton Pannekoek Institute for Astronomy, University of Amsterdam, Science Park 904, 1098 XH Amsterdam, The Netherlands}

\author[0000-0002-6092-8295]{Josh Walawender}
\affiliation{W.\ M.\ Keck Observatory, 65-1120 Mamalahoa Hwy, Kamuela, HI 96743, USA}

\author[0000-0002-6937-9034]{Sharon X.~Wang}
\affiliation{Department of Astronomy, Tsinghua University, Beijing 100084, People's Republic of China}

\author[0000-0002-3725-3058]{Lauren M. Weiss}
\affiliation{Department of Physics and Astronomy, University of Notre Dame, Notre Dame, IN 46556, USA}

\author[0000-0002-4037-3114]{Sherry Yeh}
\affiliation{W.\ M.\ Keck Observatory, 65-1120 Mamalahoa Hwy, Kamuela, HI 96743, USA}



\begin{abstract}

We introduce the OATMEAL survey, an effort to measure the obliquities of stars with transiting brown dwarf companions. We observed a transit of the close-in ($P_{\rm orb} = 1.74 \,$ days) brown dwarf GPX-1 b using the Keck Planet Finder (KPF) spectrograph to measure the sky-projected angle between its orbital axis and the spin axis of its early F-type host star ($\lambda$). We measured $\lambda = 6.9 \pm 10.0 ^\circ$, suggesting an orbit that is prograde and well aligned with the stellar equator. Hot Jupiters around early F stars are frequently found to have highly misaligned orbits, with polar and retrograde orbits being commonplace. It has been theorized that these misalignments stem from dynamical interactions, such as von Zeipel-Kozai-Lidov cycles, and are retained over long timescales due to weak tidal dissipation in stars with radiative envelopes. By comparing GPX-1 to similar systems under the frameworks of different tidal evolution theories, we argued that the rate of tidal dissipation is too slow to have re-aligned the system. This suggests that GPX-1 may have arrived at its close-in orbit via coplanar high-eccentricity migration or migration through an aligned protoplanetary disk. Our result for GPX-1 is one of few measurements of the obliquity of a star with a transiting brown dwarf. By enlarging the number of such measurements and comparing them with hot Jupiter systems, we will more clearly discern the differences between the mechanisms that dictate the formation and evolution of both classes of objects.

\end{abstract}



\section{Introduction} \label{sec:intro}

Historically, giant planets and brown dwarfs have been delineated based on mass, with a dividing line at the deuterium-burning limit ($13 \, M_{\rm J}$) \citep[e.g.,][]{spiegel2011deuterium}. However, some argue that it is more sensible to classify these objects based on their formation mechanisms and subsequent orbital evolution. While different formation mechanisms likely correlate with mass, there is currently no theoretical prediction or empirical evidence that $13 \, M_{\rm J}$ has any special significance in the process. For instance, formation via core accretion \citep{pollack1996coreaccretion}, which is thought to create most Jupiter-size planets, is possibly capable of forming objects 10--20 times more massive than Jupiter. Formation via gravitational instability \citep{boss1997grav}, which is believed to form very massive giant planets and low-mass stars efficiently, may be able to form objects as low-mass as $2 \, M_{\rm J}$ \citep[e.g.,][]{matsuo2007formation}. Post-formation evolution mechanisms, such as migration through the protoplanetary disk and high-eccentricity migration, are also believed to depend on companion mass \citep[e.g.,][]{dawson2018hotjupiters}. We can infer these histories by statistically characterizing the occurrence rates and orbital properties of giant planets and brown dwarfs.

The first studies of giant planet and brown dwarf demographics came from radial velocity surveys of Sun-like stars. Early analyses of their occurrence rates found a notable dearth of ``brown-dwarf-mass'' (hereafter $13 - 80 \, M_{\rm J}$, for the sake of convenience) objects compared to ``planetary-mass'' (hereafter $< 13 \, M_{\rm J}$) and ``stellar-mass'' (hereafter $> 80 \, M_{\rm J}$) objects at orbital separations less than $\sim 1$~AU \citep[e.g.,][]{grether2006desert}. This dearth, referred to by some as the brown dwarf ``desert,'' suggests that brown dwarfs are either unable to form close to Sun-like stars as efficiently as giant planets, or that they are rarely transported inward from their birthplaces. \citet{ma2014statistical} examined a sample of brown dwarfs detected via transit and radial velocity to explore the orbital properties of those in the desert, finding that those with masses below $42.5 \, M_{\rm J}$ have orbital eccentricities resembling giant planets and those with masses above about $42.5 \, M_{\rm J}$ have orbital eccentricities resembling stellar companions, potentially suggesting that this mass divides planet-like and star-like formations. \citet{schlaufman2018evidence} used a similar sample to inspect how the occurrences of giant planets and brown dwarfs depend on host star metallicity, finding evidence that objects with masses $< 4 \, M_{\rm J}$ tend to orbit metal-rich stars and more massive objects tend to orbit stars with a wide range of metallicities, hinting that the lower-mass objects preferentially form via core accretion.

Direct imaging surveys have also provided insight into the differences between giant planet and brown dwarf formation. For instance, \citet{nielsen2019gemini} showed that for orbital separations of $10-100$~AU, the occurrence rate of brown dwarfs is lower than that of giant planets by a factor of $\sim 10$, suggesting that the efficiencies of brown dwarf and giant planet formation mechanisms differ dramatically. \citet{bowler2020eccentricities, bowler2023obliquties} showed that wide-separation brown dwarfs tend to have orbits that are more eccentric and more misaligned with the equators of their host stars compared to wide-separation giant planets.\footnote{Although we note that the analysis of orbital eccentricities has been shown to be highly sensitive to the choice of priors \citep[e.g.,][]{do2023priors, nagpal2023priors}.} \citet{bryan2020obliquity, bryan2021obliquity} showed through measurements of the stellar spin axes, inclinations of the orbit normals, and brown dwarf spin axes of multiple systems that both the spins and orbits of wide-separation brown dwarfs are often misaligned with the spins of their stars. In general, these observations support the notion that wide-separation brown dwarfs frequently form via fragmentation of the protoplanetary disk or molecular cloud core \citep[e.g.,][]{durisen2007formation, bate2010alignment, bate2012formation, offner2016misalignment}. 

Analogous experiments can be carried out for transiting giant planets and brown dwarfs in close-in orbits. Because these close-in objects are thought to form farther from their stars than the locations in which they are observed today, their occurrence rates and orbits are likely linked more closely to their orbital evolutions than their formation \citep{dawson2018hotjupiters}. Several calculations of the hot Jupiter occurrence rate around Sun-like stars have been conducted using data from the {\it Kepler} \citep{borucki2010kepler} and {\it K2} \citep{howell2014k2} missions, revealing a frequency of $\sim 1 \%$ \citep[e.g.,][]{howard2012kepler}. More recently, {\it TESS} \citep{ricker2015tess} has enabled investigations of hot Jupiter occurrence rate as a function of stellar mass. These studies have hinted that these planets may be less common around A-type and M-type stars than around G-type stars \citep{zhou2019hotjupiter, beleznay2022hotjupiter, bryant2023hotjupiter, gan2023hotjupiter}, although the underlying cause of this trend is unknown. Similarly reliable occurrence rate calculations have yet to be performed for transiting brown dwarfs because relatively few have been found.

\begin{figure*}[t!]
  \centering
    \includegraphics[width=0.49\textwidth]{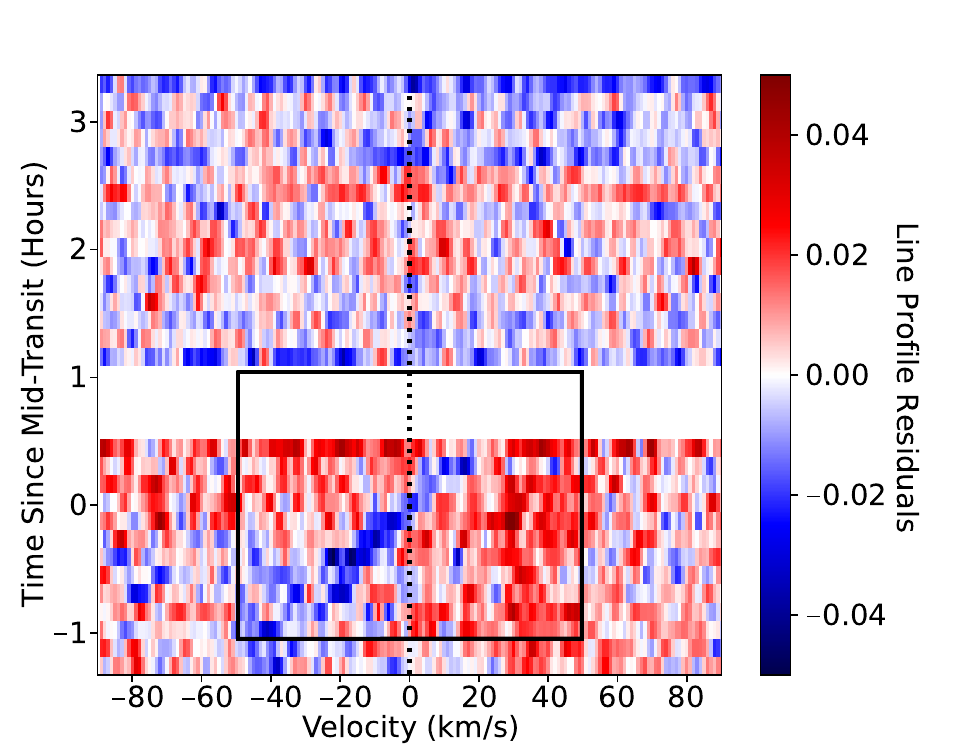}
    \includegraphics[width=0.49\textwidth]{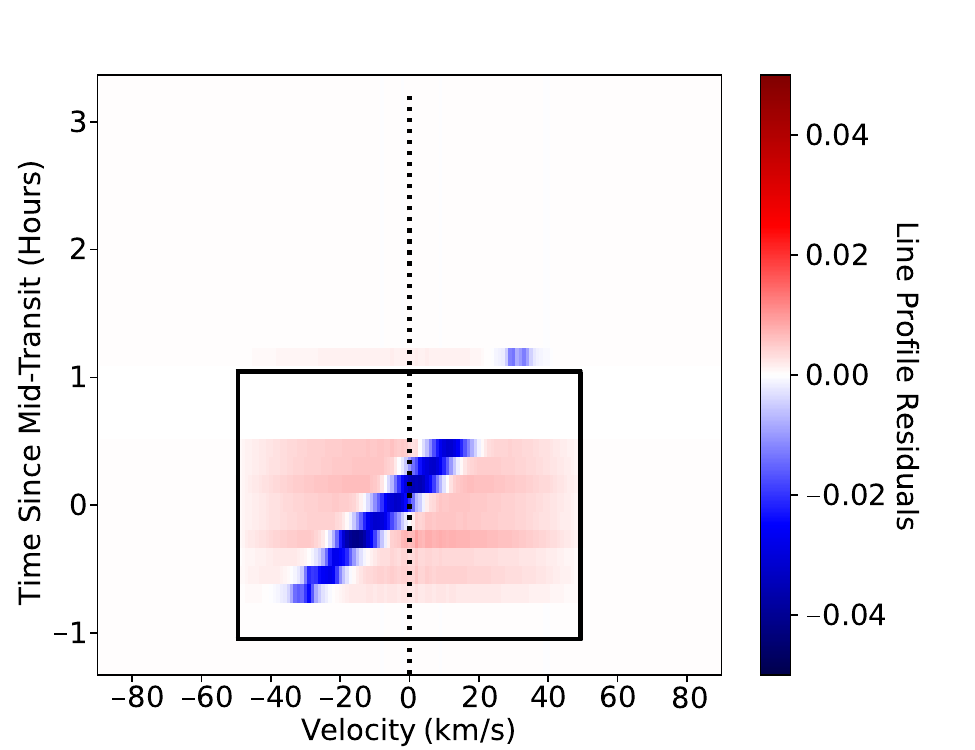}
    \caption{\emph{Left:} Residuals of the stellar line profile across the transit of GPX-1~b. The dotted lines indicate the confines of the transit and the $v \sin{i_\star}$ of the star. The transit occurs between $-1$ and $1$ hours from mid-transit. The diagonal blue line is visible in-transit is the Doppler shadow of the brown dwarf. The gap in the data near transit egress was due to a guiding malfunction that temporarily prevented data acquisition. \emph{Right:} The best-fit model of the line profile residuals, which are consistent with a sky-projected stellar obliquity of $\lambda = 6.9 \pm 1.7 ^\circ$. However, we ultimately adopted $\lambda = 6.9 \pm 10.0 ^\circ$ to account for systematic uncertainty in the model.}
    \label{fig:dopplershadow}
\end{figure*}

Stellar obliquity measurements that leverage the Rossiter-McLaughlin effect \citep{mclaughlin1924rm, rossiter1924rm} and the Doppler shadow technique \citep{colliercameron2010dopplershadow} have also provided a window into the dynamical histories of hot Jupiters. \citet{winn2010obliquity} first identified a trend in the obliquities of stars with hot Jupiters, noting that stars with effective temperatures above the Kraft break ($T_{\rm eff} \gtrsim 6250$~K) tend to have spin axes that are misaligned with the orbital axes of their planets, whereas cooler stars generally exhibit spin-orbit alignment. This finding led to two key hypotheses: (1) the processes that transport hot Jupiters from their formation locations to their current close-in orbits often leave them in orbits that are misaligned with the equators of their host stars, and (2) the ability of a star to tidally realign with the orbit of a hot Jupiter depends on $T_{\rm eff}$. In relation to the former, several mechanisms have been proposed that are capable of tilting the orbit of a hot Jupiter, although the data seems most consistent with scattering or secular interactions with a massive outer companion followed by high-eccentricity migration \citep{dawson2018hotjupiters, rice2022obliquities}. In relation to the latter, it is believed that stars with radiative envelopes ($T_{\rm eff}\gtrsim 6{,}250$\,K) are far less efficient at damping tides excited by their close-in planets than cool stars, causing a drastic discrepancy in the tidal realignment timescale and creating the dichotomy in stellar obliquities observed today.

Stellar obliquity measurements for transiting brown dwarf systems are much less common than those for transiting planet systems, largely due to the rarity of transiting brown dwarfs. However, this limitation is being lifted thanks to {\it TESS}, which has significantly increased the number of known transiting brown dwarfs around relatively bright ($V < 13$) stars \citep[e.g.,][]{carmichael2020bd, carmichael2021bd, carmichael2022bd, grieves2021bd, psaridi2022bd, lin2023bd, vowell2023bd}. These discoveries provide the exciting opportunity to investigate the stellar obliquity distribution of stars with close-orbiting brown dwarfs and compare it to that of hot Jupiters. If close-in brown dwarfs tend to have orbits that are aligned with the equators of their host stars -- especially when the stars are hot or the brown dwarfs orbit their stars at distances large enough for tides to be negligible -- it would suggest that brown dwarfs spiral inwards via disk-driven migration \citep[e.g.,][]{baruteau2014diskmigration, tokovinin2020migration} or co-planar high-eccentricity migration \citep[e.g.,][]{petrovich2015coplanar}. If these systems instead have a wide range of stellar obliquities, it would point to non-co-planar forms of secular excitation followed by high-eccentricity migration \citep[e.g.,][]{fabrycky2007kozai} or dynamical capture \citep[e.g.,][]{dorval2017binary} as the primary migration mechanisms.

As of today, obliquities have been published for only five of the roughly fifty stars known to have transiting brown dwarfs: CoRoT-3 \citep{triaud2009corot3}, KELT-1 \citep{siverd2012kelt1}, WASP-30 \citep{triaud2013wasp30}, HATS-70 \citep{zhou2019hats70}, and TOI-2533 \citep{ferreira2024toi2533}. Of these four, three orbit stars hotter than the Kraft break (the exceptions being WASP-30 and TOI-2533), and they are all consistent with good alignment (i.e., all have sky-projected stellar obliquities $\lesssim 45^\circ$). These three results hint that transiting brown dwarf systems may generally have lower stellar obliquities than transiting hot Jupiter systems, but the sample size is too small to reliably probe the underlying obliquity distribution. Here, we introduce the OATMEAL (Orbital Architectures of Transiting Massive Exoplanets And Low-mass stars)  survey, a coordinated effort to increase the number of transiting brown dwarf systems with stellar obliquity measurements. We report the first result of this survey: the stellar obliquity of the early F-type star GPX-1, which hosts a $19.7 \pm 1.6 \, M_{\rm J}$ brown dwarf with a 1.74~day orbital period \citep{benni2021gpx1}.

This paper is structured as follows. In Section~\ref{sec:observations}, we describe our observations collected with the Keck Planet Finder (KPF) spectrograph and the related data reduction. In Section~\ref{sec:results}, we outline our analysis of the data and report the stellar obliquity. In Section~\ref{sec:discussion}, we examine the possibility that the system could have undergone spin-orbit realignment and discuss plausible formation and migration scenario for the brown dwarf. Lastly, in Section~\ref{sec:conclusion}, we provide concluding remarks.

\section{Observations} \label{sec:observations}

We observed a transit of GPX-1 b on 25 September 2023 UT using the KPF spectrograph on the 10-meter Keck-I telescope \citep{gibson2016kpf, gibson2018kpf, gibson2020kpf}. We took 29 500-second exposures between 10:28 and 15:10 UT, achieving signal-to-noise ratios (SNRs) between 80 and 90 within 500--600 nm after stacking the three slices of the science fiber. These observations spanned of the brown dwarf, which took place between 11:05 UT and 13:10 UT, and approximately 2 hours of post-transit baseline. Every 60 minutes, we acquired a calibration exposure using a Fabry–Pérot etalon in order to track and correct for intranight instrumental drift. The data was reduced using the publicly available KPF pipeline,\footnote{\url{https://github.com/Keck-DataReductionPipelines/KPF-Pipeline}} which produces wavelength-calibrated and barycentric-corrected spectra for each order on the CCD. 

We note that there is a gap in the data between 12:18 and 12:52 UT, which was due to a software malfunction related to the tip-tilt mirror used for guiding. However, as we show below, this gap in the data did not significantly impact our ability to measure the stellar obliquity of the system.

\section{Analysis and Results} \label{sec:results}

The goal of our observations were to measure the sky-projected obliquity ($\lambda$) of the star GPX-1, or the sky-projected angle between its spin axis and the orbital axis of its transiting brown dwarf companion. For each spectrum, we calculated the stellar line profile using the least-squares deconvolution (LSD) method \citep{donati1997lsd}. Calculation of the profile via LSD required a binary mask of stellar lines, which we obtained from version 3 of the Vienna Atomic Line Database (VALD3; \citealt{ryabchikova2015vald3}) for a star with properties matching those of GPX-1 reported in \cite{benni2021gpx1}. Next, we narrowed down this line list by inspecting the collected spectra for lines that are deep enough to see visually, are relatively free from blends arising from rotational broadening, and are uncontaminated by strong telluric features (e.g., the molecular oxygen A- and B-bands). These 184 lines provided the best characterization of the stellar line profile.


\begin{figure*}[t!]
  \centering
    \includegraphics[width=\textwidth]{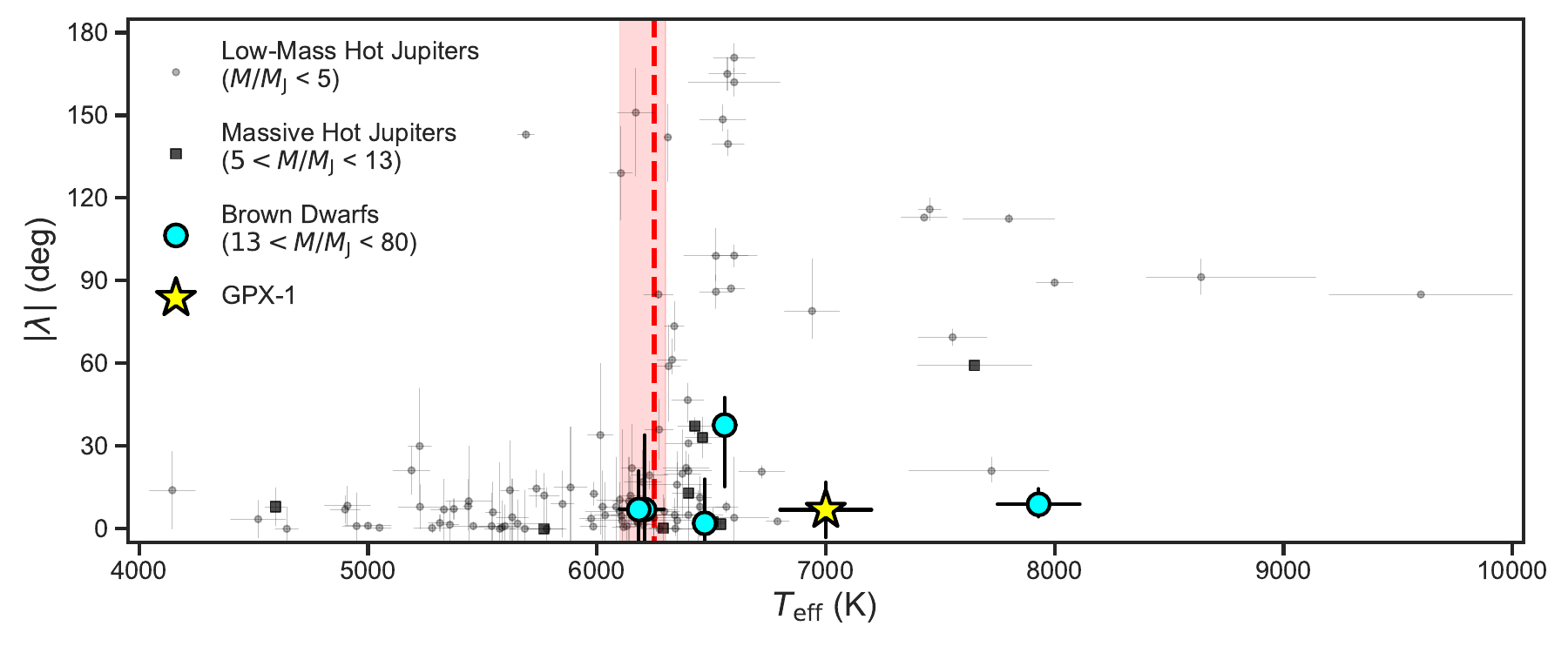}
    \caption{Sky-projected stellar obliquity ($| \lambda |$) versus stellar effective temperature ($T_{\rm eff}$) for systems with low-mass hot Jupiters (light gray dots), massive hot Jupiters (dark gray squares), and close-in brown dwarfs (blue circles). GPX-1 is the yellow star. The dashed vertical red line and surrounding shaded area is the Kraft break regime (we adopt $T_{\rm Kraft} = 6250$~K). The data for the hot Jupiters were taken from Tables A1 and A2 of \citet{albrecht2022obliquity}, which were originally sourced from TEPCat \citep{southworth2011tepcat}. The four previously characterized transiting brown dwarf systems are WASP-30 ($T_{\rm eff} = 6208 \pm 85$~K, $| \lambda | = 7^\circ$$^{+19}_{-27}$; \citealt{triaud2013wasp30}), KELT-1 ($T_{\rm eff} = 6471 \pm 50$~K, $| \lambda | = 2 \pm 16^\circ$; \citealt{siverd2012kelt1}), CoRoT-3 ($T_{\rm eff} = 6558 \pm 50$~K, $| \lambda | = 38^\circ$$_{-22}^{+10}$; \citealt{triaud2009corot3}), HATS-70 ($T_{\rm eff} = 7930 \pm 180$~K, $| \lambda | = 8.9^\circ$$^{+5.6}_{-4.5}$; \citealt{zhou2019hats70}), and TOI-2533 ($T_{\rm eff} = 6183^{+16}_{-84}$~K, $| \lambda | = 7 \pm 14^\circ$; \citealt{ferreira2024toi2533}). Note that WASP-30 and TOI-2533 are overlapping. GPX-1 contributes to growing trend for hot stars with close-in brown dwarfs and massive hot Jupiters to have lower values of $\lambda$ than hot stars with low-mass hot Jupiters, although the sample size is still too small to claim that these populations differ in a statistically significant way.}
    \label{fig:obliquities}
\end{figure*}

We examined the Doppler Shadow of GPX-1 using the method described in \citet{Dai2020}. In short, we subtracted the average out-of-transit line profile from the in-transit line profile. Figure \ref{fig:dopplershadow} shows the line profile residuals as a function of time and stellar-centric velocity. The Doppler shadow of GPX-1 is the blue diagonal feature in the lower half of the diagram. The flux deficit in the line profile moves from the blueshifted to the redshifted limb of the host star. This is expected if the BD is on an aligned orbit around the host star. The star is assumed to have solid-body rotation, while GPX-1 is assumed to have a circular orbit \citep{benni2021gpx1} with elements fixed at the median values reported by \citep{benni2021gpx1}. The only exceptions are the impact parameter and sky-projected obliquity of the BD which were allowed to vary freely in our Doppler shadow model. In the model, we synthesized line profiles by pixelating the host star. Each pixel is assigned a line-of-sight rotational velocity, limb darkening, and macroturbulence, all of which are nuisance parameters that are marginalized in posterior sampling. Our posterior sampling is performed with nested sampling code using {\sc Dynesty} \citep{Speagle} using the default sampling settings. We found $\lambda = 6.9 \pm 1.7 ^\circ$ and $v \sin{i_\star} = 50.77^{+2.27}_{-2.62}$~km~s$^{-1}$, suggesting a prograde orbit that is well-aligned with the stellar equator. We note that this uncertainty on $\lambda$ is likely underestimated as it does not include any systematic errors associated with various above-mentioned assumptions and additional effects not included in the model, such as instrumental line profile variations. We predict the true uncertainty to be 3-6 times higher than that provided by our fit. To err on the conservative side, we instead report an obliquity of $\lambda = 6.9 \pm 10.0 ^\circ$. Lastly, we stress that because we do not know the inclination of the stellar spin axis, the true 3D stellar obliquity may be higher. GPX-1 likely has a fully radiative exterior and a spotless surface, making it challenging to determine the rotation period (and therefore inclination) of the star. Indeed, we found no evidence of starspot modulation in the {\it TESS} data.

\section{Discussion} \label{sec:discussion}

We display GPX-1 in the context of other transiting hot Jupiter and brown dwarf systems with $\lambda$ measurements in the top panel of Figure \ref{fig:obliquities}. Like other stars with massive hot Jupiters and close-in brown dwarfs for which $\lambda$ is known, GPX-1 is consistent with spin-orbit alignment. This measurement contrasts with those of hot-Jupiter hosts with $T_{\rm eff} \geq 7000$~K, for which the lowest measured $\lambda$ is $21^\circ$$_{-4}^{+5}$ (HAT-P-69; \citealt{zhou2019hotjupiter}).

The observation of good alignment raises the question of whether GPX-1~b arrived at its close-in orbit on an already-aligned orbit or if the system was able to achieve spin-orbit realignment on a shorter timescale than systems with hot Jupiters. On one hand, GPX-1~b is more massive ($m = 19.7 \pm 1.6 \, M_{\rm J}$; \citealt{benni2021gpx1}) than most hot Jupiters, which should correspond to a faster tidal realignment. On the other hand, GPX-1 is hot enough to have a fully radiative outer envelope and is estimated to be relatively young (${\rm Age} = 0.27^{+0.09}_{-0.15}$~Gyr; \citealt{benni2021gpx1}), meaning that tidal damping in the star may not have been efficient enough to realign the system to its current orientation. Here, we consider whether or not enough time has transpired for the GPX-1 to have realigned tidally, assuming the reported age is accurate.

\begin{figure*}[t!]
  \centering
    \includegraphics[width=\textwidth]{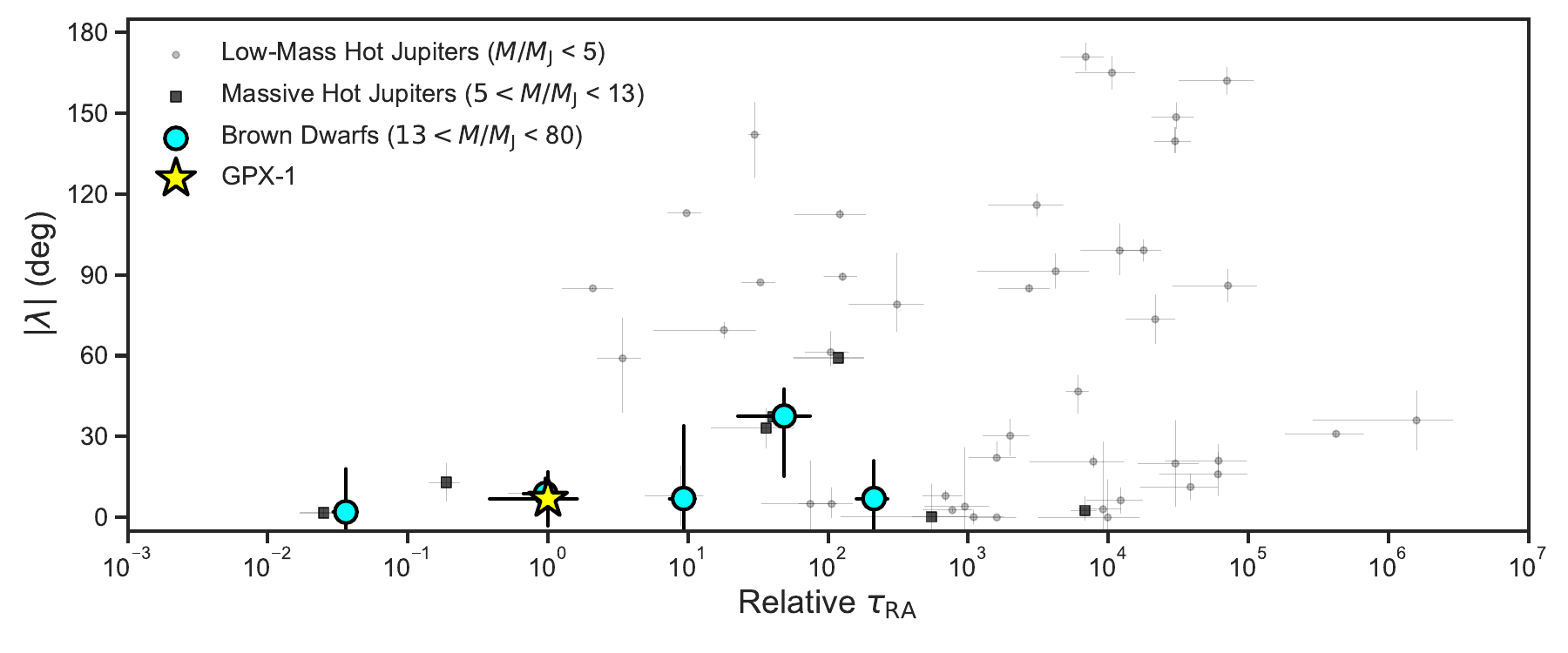}
    \caption{Relative realignment timescales using the scaling relation from \citet{zahn1975dynamicaltide, zahn1977tidal} (i.e., Equation \ref{eq:RA}), normalized to that of GPX-1. Only hot Jupiters and brown dwarfs orbiting hot ($T_{\rm eff} > 6250$ K) stars are included, with the exception of the transiting brown dwarf system WASP-30, which straddles the Kraft break. From lowest to highest $\tau_{\rm RA}$, the brown-dwarf-hosting stars are KELT-1, HATS-70 (partially obstructed by GPX-1), WASP-30, CoRoT-3, and TOI-2533. We note that the timescales for WASP-30 and TOI-2533 may not be fully representative of realignment in those systems because they likely have convective envelopes, but we include them for the sake of completeness. GPX-1 has a predicted timescale roughly $100 \times$ longer than KELT-1, which is believed to have undergone significant tidal evolution \citep{siverd2012kelt1}, and roughly equal to HATS-70 (partially hidden underneath GPX-1 in the figure), which is thought to have undergone relatively little tidal evolution \citep{zhou2019hats70}.}
    \label{fig:realignment}
\end{figure*}


A common method of determining if a system has previously undergone significant tidal evolution is by assessing whether the system has reached tidal equilibrium, which is characterized by a circular orbit, spin-orbit alignment, and synchronous rotation \citep{hut1980equil}. In general, a tidal equilibrium state is achieved if the total angular momentum of the system $L_{\rm tot}$ is larger than some critical value $L_{\rm crit}$, where
\begin{equation}
    L_{\rm tot} = L_{\rm orb} + (\alpha_\star M_\star R_\star^2 + \alpha m r^2) n
\end{equation}
and
\begin{equation}
    L_{\rm crit} = 4 \left[ \frac{G^2}{27} \frac{M_\star^3 m^3}{M_\star + m} (\alpha_\star M_\star R_\star^2 + \alpha m r^2) \right]^{1/4}.
\end{equation}
Here, $L_{\rm orb} = \frac{M_\star m}{\sqrt{M_\star + m}} \sqrt{G a (1 - e^2)}$ is the orbital angular momentum, $G$ is the gravitational constant, $a$ is the semi-major axis, $e$ is the orbital eccentricity, $M_\star$ is the mass of the host star, $R_\star$ is the radius of the host star, $m$ is the mass of the companion, $r$ is the radius of the companion, $\alpha_\star = 0.06$ is the square of the radius of stellar gyration, $\alpha = 0.26$ is the square of the radius of companion gyration, and $n$ is the mean motion \citep{matsumura2010tidal}. We adopted parameters for the GPX-1 system from \citet{benni2021gpx1}: $a = 0.0338 \pm 0.0003$~AU, $e = 0$, $M_\star = 1.68 \pm 0.10 \, M_\odot$, $R_\star = 1.56 \pm 0.10 \, R_\odot$, $m = 19.7 \pm 1.6 \, M_{\rm J}$, $r = 1.47 \pm 0.10 \, R_{\rm J}$. Assuming $i_\star = 90^\circ$, we calculated $L_{\rm tot}/L_{\rm crit} = 1.032 \pm 0.024$ for GPX-1. This represents the minimum possible ratio, as lower values of $i_\star$ correspond to faster stellar rotation speeds and therefore higher values of $L_{\rm tot}$. Assuming an isotropic distribution of possible stellar spin vectors, we found the possible system orientations divided roughly evenly between ``Darwin stable'' (at higher $i_\star$; corresponding to a system where the orbit of the companion no longer evolves tidally) and ``Darwin unstable'' (at lower $i_\star$; corresponding to a system where the orbit of the companion will eventually decay to within the Roche limit of the star). By comparison, we calculated minimum values of $L_{\rm tot}/L_{\rm crit} = 1.198 \pm 0.034$ for CoRoT-3 \citep{triaud2009corot3}, $L_{\rm tot}/L_{\rm crit} = 0.995 \pm 0.011$ for KELT-1 \citep{siverd2012kelt1}, $L_{\rm tot}/L_{\rm crit} = 1.51 \pm 0.09$ for WASP-30 \citep{triaud2013wasp30}, $L_{\rm tot}/L_{\rm crit} = 0.951 \pm 0.014$ for HATS-70 \citep{zhou2019hats70}, and $L_{\rm tot}/L_{\rm crit} = 2.09 \pm 0.05$ for TOI-2533 \citep{ferreira2024toi2533}. Taken at face value, these numbers suggest that many close-in brown dwarfs have reached, or are close to reaching, a state of tidal equilibrium \citep[e.g.,][]{husnoo2012tides}. However, we also see that these ratios do not correlate with measured spin-orbit angles. For example, CoRoT-3 should be in tidal equilibrium but was measured to have a relatively high degree of misalignment with $\lambda = 38^\circ$$^{+10}_{-22}$, indicating that tides have not yet fully realigned the spin of the star with the orbit of its companion.\footnote{Although we note that this measurement has relatively large error bars and that more precise observations could plausibly yield a much smaller $\lambda$.} Additionally, over half of hot stars with known misaligned ($\lambda > 30^\circ$) hot Jupiters also have $L_{\rm tot}/L_{\rm crit} > 1$. This tidal equilibrium metric therefore does not appear to be a reliable predictor of spin-orbit realignment due to tidal damping in hot stars. 

Alternatively, we can estimate the timescale of spin-orbit realignment predicted by different tidal damping frameworks. Multiple theories have been proposed to explain the distribution of stellar obliquities for close-in exoplanet systems (we refer the reader to Section~4 of \citealt{albrecht2022obliquity} for a summary); we explore a few of these theories here.

Firstly, we acknowledge that previous studies of spin-orbit realignment have often utilized classical equilibrium-tide theory \citep[e.g.,][]{winn2010obliquity, dawson2014equil, rice2022obliquities}. However, as is discussed and demonstrated in \citet{albrecht2022obliquity}, equilibrium tides alone cannot reproduce the observed distribution of close-in companions around hot stars. One major issue with equilibrium-tide theory is that it predicts the obliquity realignment timescale to be comparable to the orbital decay timescale of the close-in companion. If this were the case, most hot Jupiters around cool stars that have tidally realigned would have also been engulfed, which is inconsistent with the data. In addition, realignment via equilibrium tides is typically thought to be dependent on interactions between the orbit of the close-in companion and the convective envelope of the star \citep[e.g.,][]{hut1980equil}, which hot stars like GPX-1 do not have. We therefore turn to other frameworks for estimating the realignment timescale.


\citet{zahn1975dynamicaltide, zahn1977tidal} calculated the synchronization times of F and A binary stars, assuming internal gravity waves tidally excited at the base of the star's envelope dissipate by radiative diffusion. \citet{albrecht2012obliquity} used this synchronization time as a proxy for the obliquity realignment timescale for substellar companions:
\begin{equation}\label{eq:RA}
    \tau_{\rm RA} \propto Q_\star \left(\frac{m}{M_\star}\right)^{-2} \left(1 + \frac{m}{M_\star}\right)^{-5/6} \left( \frac{a}{R_\star} \right)^{17/2},
\end{equation}
where $Q_\star$ is the stellar tidal quality factor. An honest calculation of $Q_\star$ is complex (see e.g. \citealt{zahn1975dynamicaltide, Hurley+(2002), Kushnir+(2017), SuLai(2022)}). Instead, we compare the relative value of $\tau_{\rm RA}$ for stars with $T_{\rm eff} > 6250$ K, assuming a common value of $Q_\star$, in order to search for trends with $\lambda$ and put GPX-1 into context with similar systems. The results of this calculation are shown in Figure~\ref{fig:realignment}.

\begin{figure}[t!]
  \centering
    \includegraphics[width=0.48\textwidth]{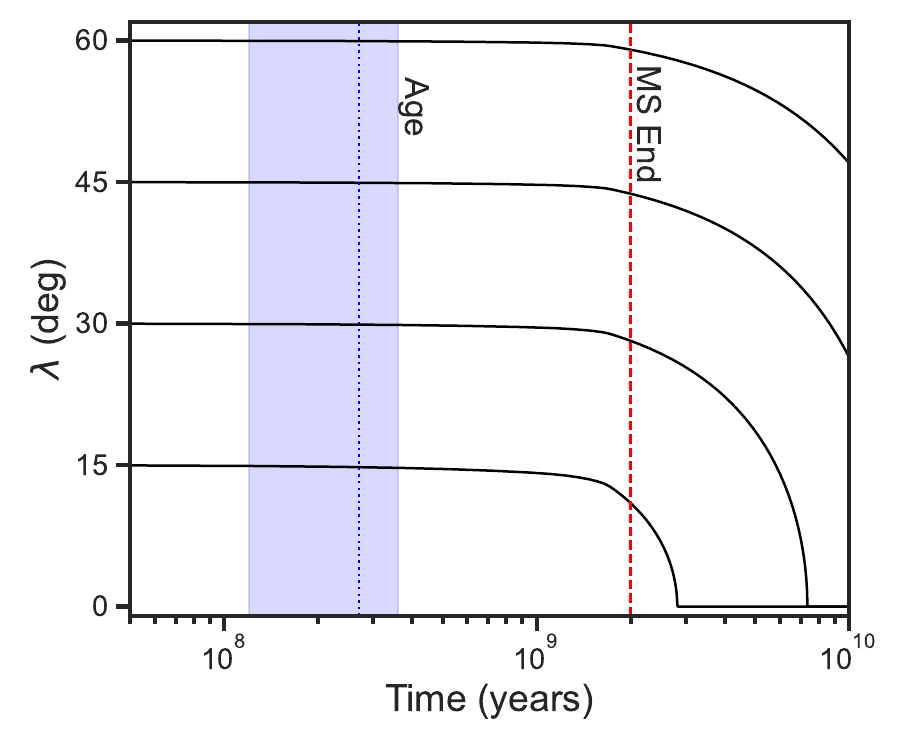}
    \caption{Simulated evolution of $\lambda$ in the GPX-1 system using the resonance locking framework from \citet{zanazzi2024damping} for four different initial misaligned values of $\lambda$. These simulations assume that the orbit of the brown dwarf begins close-in and circular (i.e., we ignore the tidal evolution of semi-major axis and eccentricity). The red dashed line is the approximate main sequence lifetime for GPX-1 and the blue dotted line is the reported age of GPX-1 (with $1\sigma$ uncertainties; \citealt{benni2021gpx1}). The model predicts that GPX-1 has undergone negligible $\lambda$ evolution during its lifetime, primarily because the $g$-mode frequencies of GPX-1-like stars only evolve significantly after the first Gyr. This supports the notion that GPX-1~b arrived at its short-period orbit in an already-aligned state. We note that this simulation neglects differences in internal stellar structure during the pre-main-sequence and post-main-sequence phases, which are likely more conducive to rapid spin-orbit realignment.}
    \label{fig:resonance_locking}
\end{figure}

According to this relative timescale, GPX-1 should realign quicker than most other hot Jupiters and brown dwarfs. The brown dwarf with a smaller $\tau_{\rm RA}$ than GPX-1 is KELT-1, which \citet{siverd2012kelt1} argued has likely synchronized with the orbit of the close-in companion due to tidal interactions. This argument was made by comparing the approximate equatorial velocity of the star, calculated using $v \sin{i_\star}$ and assuming $i_\star = 90^\circ$, with the orbital velocity of the companion. \citet{zhou2019hats70} used the same reasoning to argue that HATS-70 is {\it not} synchronized with the orbit of its brown dwarf companion and therefore has {\it not} undergone significant tidal evolution. We can say the same for GPX-1 and its companion, which have an equatorial velocity of $50.77^{+2.27}_{-2.62}$ km s$^{-1}$ (assuming $i_\star = 90^\circ$) and an orbital velocity of $211 \pm 2$ km s$^{-1}$, respectively.\footnote{Of course, one could satisfy the synchronization condition with a smaller value of $i_\star$, but a smaller $i_\star$ would also result in a higher 3D stellar obliquity. There is therefore no situation in which both synchronization and spin-orbit alignment can be achieved.} Indeed, GPX-1 and HATS-70 have roughly equal values of $\tau_{\rm RA}$. This suggests that either (1) the GPX-1 and HATS-70 systems have not yet undergone significant tidal evolution, or (2) tidal realignment occurs significantly faster than tidal synchronization in hot stars.

Another popular driver of tidal evolution is resonance locking, the process by which the orbit of the close-in companion couples with the gravity mode of the star \citep{savonije2008resonancelock, fuller2017resonancelock, ma2021resonancelock, zanazzi2021resonancelock}. Recently, \citet{zanazzi2024damping} showed that resonance locking can explain the distribution of stellar obliquities as a function of stellar $T_{\rm eff}$. In short, this mechanism can explain the low obliquities of cool stars and the high obliquities of hot stars because $g$-mode frequencies increase substantially over the main sequence phase in stars with radiative cores, whereas $g$-mode frequencies remain relatively constant in stars with convective cores. For a system in resonance lock, this frequency evolution drives strong tidal evolution. Under this framework, the evolution of the 3D obliquity ($\psi$) goes like
\begin{equation}\label{eq:RL}
    \frac{d\psi}{dt} \simeq \left( \frac{J_{\rm orb}}{J_\star} + \cos{\psi} \right) \frac{1}{3 t_{\rm ev} \sin{\psi}}
\end{equation}
where $t_{\rm ev}$ is the timescale of $g$-mode frequency evolution, \mbox{$J_{\rm orb} = m a^2 \Omega$}, \mbox{$J_\star \sim 0.04 M_\star R_\star^2 \Omega_\star$} for F and A stars, \mbox{$\Omega = \sqrt{G M_\star / a^3}$}, and $\Omega_\star$ is the stellar spin frequency. Here, we estimate $t_{\rm ev}$ for the star as a function of age by interpolating the data in Figure 6 of \citet{zanazzi2024damping} and we assume $\psi \approx \lambda$ (i.e., $i_\star \approx 90^\circ$). The evolution of $\lambda$ for four different initial misaligned orientations is shown in Figure~\ref{fig:resonance_locking}. The model predicts that for an initial misaligned orbit, very little $\lambda$ evolution would have occurred in the GPX-1 system thus far. In other words, realignment via resonance locking cannot explain the low obliquity of GPX-1 that we observe today.


These interpretations are compatible with a primordially aligned system produced by inward migration via disk migration \citep{baruteau2014diskmigration, tokovinin2020migration} or co-planar high-eccentricity migration \citep{petrovich2015coplanar}. We note that the latter preferentially occurs in the presence of a more massive outer companion responsible for the excitation of the orbital eccentricity of the inner companion. This would necessitate the existence of an additional undetected brown dwarf or star in the outer system, which may be detectable in upcoming \textit{Gaia} data releases \citep[e.g.,][]{gaia2022}. In either case, these migration mechanisms require the brown dwarf to have formed near the midplane of the protoplanetary disk, pointing to core accretion or disk fragmentation as the most likely formation mechanisms for GPX-1~b.

\section{Conclusion} \label{sec:conclusion}

We reported a sky-projected obliquity of $\lambda = 6.9 \pm 10.0 ^\circ$ for the early F-type star GPX-1, which hosts a close-in transiting brown dwarf. This measurement suggests that the orbit of the brown dwarf is prograde and that the spin of the star and the orbit of the planet are well aligned. This orientation is consistent with other known transiting brown dwarf systems, but is unlike the many hot Jupiters orbiting hot stars that frequently have polar and retrograde orbits. We argued that, if the relatively young reported age {\rm for} the system (${\rm Age} = 0.27^{+0.09}_{-0.15}$~Gyr; \citealt{benni2021gpx1}) is correct, the brown dwarf is most likely primordially aligned due to inefficient tidal damping in hot stars. We encourage the measurement of stellar obliquity for more transiting brown dwarf systems, which would allow for a statistically robust characterization of the underlying obliquity distribution and provide novel clues for the origins of elusive close-in brown dwarfs.

\begin{acknowledgements}

We thank Luke Bouma for his useful feedback and suggestions.

The data presented herein were obtained at the W. M. Keck Observatory, which is operated as a scientific partnership among the California Institute of Technology, the University of California, and the National Aeronautics and Space Administration. The Observatory was made possible by the generous financial support of the W. M. Keck Foundation. Keck Observatory occupies the summit of Maunakea, a place of significant ecological, cultural, and spiritual importance within the indigenous Hawaiian community. We understand and embrace our accountability to Maunakea and the indigenous Hawaiian community, and commit to our role in long-term mutual stewardship. We are most fortunate to have the opportunity to conduct observations from Maunakea.

We gratefully acknowledge the efforts and dedication of the Keck Observatory staff for support of KPF and remote observing.

SG is supported by an NSF Astronomy and Astrophysics Postdoctoral Fellowship under award AST-2303922.

This research was carried out, in part, at the Jet Propulsion Laboratory and the California Institute of Technology under a contract with the National Aeronautics and Space Administration and funded through the President’s and Director’s Research \& Development Fund Program.

\end{acknowledgements}
 
\bibliography{bibliography}{}

\begin{thebibliography}{}
\expandafter\ifx\csname natexlab\endcsname\relax\def\natexlab#1{#1}\fi
\providecommand{\url}[1]{\href{#1}{#1}}
\providecommand{\dodoi}[1]{doi:~\href{http://doi.org/#1}{\nolinkurl{#1}}}
\providecommand{\doeprint}[1]{\href{http://ascl.net/#1}{\nolinkurl{http://ascl.net/#1}}}
\providecommand{\doarXiv}[1]{\href{https://arxiv.org/abs/#1}{\nolinkurl{https://arxiv.org/abs/#1}}}

\bibitem[{{Albrecht} {et~al.}(2012){Albrecht}, {Winn}, {Johnson}, {Howard}, {Marcy}, {Butler}, {Arriagada}, {Crane}, {Shectman}, {Thompson}, {Hirano}, {Bakos}, \& {Hartman}}]{albrecht2012obliquity}
{Albrecht}, S., {Winn}, J.~N., {Johnson}, J.~A., {et~al.} 2012, \apj, 757, 18, \dodoi{10.1088/0004-637X/757/1/18}

\bibitem[{{Albrecht} {et~al.}(2022){Albrecht}, {Dawson}, \& {Winn}}]{albrecht2022obliquity}
{Albrecht}, S.~H., {Dawson}, R.~I., \& {Winn}, J.~N. 2022, \pasp, 134, 082001, \dodoi{10.1088/1538-3873/ac6c09}

\bibitem[{{Baruteau} {et~al.}(2014){Baruteau}, {Crida}, {Paardekooper}, {Masset}, {Guilet}, {Bitsch}, {Nelson}, {Kley}, \& {Papaloizou}}]{baruteau2014diskmigration}
{Baruteau}, C., {Crida}, A., {Paardekooper}, S.~J., {et~al.} 2014, in Protostars and Planets VI, ed. H.~{Beuther}, R.~S. {Klessen}, C.~P. {Dullemond}, \& T.~{Henning}, 667--689, \dodoi{10.2458/azu_uapress_9780816531240-ch029}

\bibitem[{{Bate}(2012)}]{bate2012formation}
{Bate}, M.~R. 2012, \mnras, 419, 3115, \dodoi{10.1111/j.1365-2966.2011.19955.x}

\bibitem[{{Bate} {et~al.}(2010){Bate}, {Lodato}, \& {Pringle}}]{bate2010alignment}
{Bate}, M.~R., {Lodato}, G., \& {Pringle}, J.~E. 2010, \mnras, 401, 1505, \dodoi{10.1111/j.1365-2966.2009.15773.x}

\bibitem[{{Beleznay} \& {Kunimoto}(2022)}]{beleznay2022hotjupiter}
{Beleznay}, M., \& {Kunimoto}, M. 2022, \mnras, 516, 75, \dodoi{10.1093/mnras/stac2179}

\bibitem[{{Benni} {et~al.}(2021){Benni}, {Burdanov}, {Krushinsky}, {Bonfanti}, {H{\'e}brard}, {Almenara}, {Dalal}, {Demangeon}, {Tsantaki}, {Pepper}, {Stassun}, {Vanderburg}, {Belinski}, {Kashaev}, {Barkaoui}, {Kim}, {Kang}, {Antonyuk}, {Dyachenko}, {Rastegaev}, {Beskakotov}, {Mitrofanova}, {Pozuelos}, {Kuznetsov}, {Popov}, {Kiefer}, {Wilson}, {Ricker}, {Vanderspek}, {Latham}, {Seager}, {Jenkins}, {Sokov}, {Sokova}, {Marchini}, {Papini}, {Salvaggio}, {Banfi}, {Ba{\c{s}}t{\"u}rk}, {Torun}, {Yal{\c{c}}{\i}nkaya}, {Ivanov}, {Valyavin}, {Jehin}, {Gillon}, {Pak{\v{s}}tien{\.{e}}}, {Hentunen}, {Shadick}, {Bretton}, {W{\"u}nsche}, {Garlitz}, {Jongen}, {Molina}, {Girardin}, {Grau Horta}, {Naves}, {Benkhaldoun}, {Joner}, {Spencer}, {Bieryla}, {Stevens}, {Jensen}, {Collins}, {Charbonneau}, {Quintana}, {Mullally}, \& {Henze}}]{benni2021gpx1}
{Benni}, P., {Burdanov}, A.~Y., {Krushinsky}, V.~V., {et~al.} 2021, \mnras, 505, 4956, \dodoi{10.1093/mnras/stab1567}

\bibitem[{{Borucki} {et~al.}(2010){Borucki}, {Koch}, {Basri}, {Batalha}, {Brown}, {Caldwell}, {Caldwell}, {Christensen-Dalsgaard}, {Cochran}, {DeVore}, {Dunham}, {Dupree}, {Gautier}, {Geary}, {Gilliland}, {Gould}, {Howell}, {Jenkins}, {Kondo}, {Latham}, {Marcy}, {Meibom}, {Kjeldsen}, {Lissauer}, {Monet}, {Morrison}, {Sasselov}, {Tarter}, {Boss}, {Brownlee}, {Owen}, {Buzasi}, {Charbonneau}, {Doyle}, {Fortney}, {Ford}, {Holman}, {Seager}, {Steffen}, {Welsh}, {Rowe}, {Anderson}, {Buchhave}, {Ciardi}, {Walkowicz}, {Sherry}, {Horch}, {Isaacson}, {Everett}, {Fischer}, {Torres}, {Johnson}, {Endl}, {MacQueen}, {Bryson}, {Dotson}, {Haas}, {Kolodziejczak}, {Van Cleve}, {Chandrasekaran}, {Twicken}, {Quintana}, {Clarke}, {Allen}, {Li}, {Wu}, {Tenenbaum}, {Verner}, {Bruhweiler}, {Barnes}, \& {Prsa}}]{borucki2010kepler}
{Borucki}, W.~J., {Koch}, D., {Basri}, G., {et~al.} 2010, Science, 327, 977, \dodoi{10.1126/science.1185402}

\bibitem[{{Boss}(1997)}]{boss1997grav}
{Boss}, A.~P. 1997, Science, 276, 1836, \dodoi{10.1126/science.276.5320.1836}

\bibitem[{{Bowler} {et~al.}(2020){Bowler}, {Blunt}, \& {Nielsen}}]{bowler2020eccentricities}
{Bowler}, B.~P., {Blunt}, S.~C., \& {Nielsen}, E.~L. 2020, \aj, 159, 63, \dodoi{10.3847/1538-3881/ab5b11}

\bibitem[{{Bowler} {et~al.}(2023){Bowler}, {Tran}, {Zhang}, {Morgan}, {Ashok}, {Blunt}, {Bryan}, {Evans}, {Franson}, {Huber}, {Nagpal}, {Wu}, \& {Zhou}}]{bowler2023obliquties}
{Bowler}, B.~P., {Tran}, Q.~H., {Zhang}, Z., {et~al.} 2023, \aj, 165, 164, \dodoi{10.3847/1538-3881/acbd34}

\bibitem[{{Bryan} {et~al.}(2021){Bryan}, {Chiang}, {Morley}, {Mace}, \& {Bowler}}]{bryan2021obliquity}
{Bryan}, M.~L., {Chiang}, E., {Morley}, C.~V., {Mace}, G.~N., \& {Bowler}, B.~P. 2021, \aj, 162, 217, \dodoi{10.3847/1538-3881/ac1bb1}

\bibitem[{{Bryan} {et~al.}(2020){Bryan}, {Chiang}, {Bowler}, {Morley}, {Millholland}, {Blunt}, {Ashok}, {Nielsen}, {Ngo}, {Mawet}, \& {Knutson}}]{bryan2020obliquity}
{Bryan}, M.~L., {Chiang}, E., {Bowler}, B.~P., {et~al.} 2020, \aj, 159, 181, \dodoi{10.3847/1538-3881/ab76c6}

\bibitem[{{Bryant} {et~al.}(2023){Bryant}, {Bayliss}, \& {Van Eylen}}]{bryant2023hotjupiter}
{Bryant}, E.~M., {Bayliss}, D., \& {Van Eylen}, V. 2023, \mnras, 521, 3663, \dodoi{10.1093/mnras/stad626}

\bibitem[{{Carmichael} {et~al.}(2020){Carmichael}, {Quinn}, {Mustill}, {Huang}, {Zhou}, {Persson}, {Nielsen}, {Collins}, {Ziegler}, {Collins}, {Rodriguez}, {Shporer}, {Brahm}, {Mann}, {Bouchy}, {Fridlund}, {Stassun}, {Hellier}, {Seidel}, {Stalport}, {Udry}, {Pepe}, {Ireland}, {{\v{Z}}erjal}, {Brice{\~n}o}, {Law}, {Jord{\'a}n}, {Espinoza}, {Henning}, {Sarkis}, \& {Latham}}]{carmichael2020bd}
{Carmichael}, T.~W., {Quinn}, S.~N., {Mustill}, A.~J., {et~al.} 2020, \aj, 160, 53, \dodoi{10.3847/1538-3881/ab9b84}

\bibitem[{{Carmichael} {et~al.}(2021){Carmichael}, {Quinn}, {Zhou}, {Grieves}, {Irwin}, {Stassun}, {Vanderburg}, {Winn}, {Bouchy}, {Brasseur}, {Brice{\~n}o}, {Caldwell}, {Charbonneau}, {Collins}, {Colon}, {Eastman}, {Fausnaugh}, {Fong}, {F{\H{u}}r{\'e}sz}, {Huang}, {Jenkins}, {Kielkopf}, {Latham}, {Law}, {Lund}, {Mann}, {Ricker}, {Rodriguez}, {Schwarz}, {Shporer}, {Tenenbaum}, {Wood}, \& {Ziegler}}]{carmichael2021bd}
{Carmichael}, T.~W., {Quinn}, S.~N., {Zhou}, G., {et~al.} 2021, \aj, 161, 97, \dodoi{10.3847/1538-3881/abd4e1}

\bibitem[{{Carmichael} {et~al.}(2022){Carmichael}, {Irwin}, {Murgas}, {Pall{\'e}}, {Stassun}, {Bartnik}, {Collins}, {de Leon}, {Esparza-Borges}, {Fedewa}, {Fong}, {Fukui}, {Jenkins}, {Kagetani}, {Latham}, {Lund}, {Mann}, {Moldovan}, {Morgan}, {Narita}, {Painter}, {Parviainen}, {Quintana}, {Ricker}, {Schulte}, {Schwarz}, {Seager}, {Sokolovsky}, {Twicken}, \& {Winn}}]{carmichael2022bd}
{Carmichael}, T.~W., {Irwin}, J.~M., {Murgas}, F., {et~al.} 2022, \mnras, 514, 4944, \dodoi{10.1093/mnras/stac1666}

\bibitem[{{Collier Cameron} {et~al.}(2010){Collier Cameron}, {Bruce}, {Miller}, {Triaud}, \& {Queloz}}]{colliercameron2010dopplershadow}
{Collier Cameron}, A., {Bruce}, V.~A., {Miller}, G.~R.~M., {Triaud}, A.~H.~M.~J., \& {Queloz}, D. 2010, \mnras, 403, 151, \dodoi{10.1111/j.1365-2966.2009.16131.x}

\bibitem[{{Dai} {et~al.}(2020){Dai}, {Roy}, {Fulton}, {Robertson}, {Hirsch}, {Isaacson}, {Albrecht}, {Mann}, {Kristiansen}, {Batalha}, {Beard}, {Behmard}, {Chontos}, {Crossfield}, {Dalba}, {Dressing}, {Giacalone}, {Hill}, {Howard}, {Huber}, {Kane}, {Kosiarek}, {Lubin}, {Mayo}, {Mocnik}, {Akana Murphy}, {Petigura}, {Rosenthal}, {Rubenzahl}, {Scarsdale}, {Weiss}, {Van Zandt}, {Ricker}, {Vanderspek}, {Latham}, {Seager}, {Winn}, {Jenkins}, {Caldwell}, {Charbonneau}, {Daylan}, {G{\"u}nther}, {Morgan}, {Quinn}, {Rose}, \& {Smith}}]{Dai2020}
{Dai}, F., {Roy}, A., {Fulton}, B., {et~al.} 2020, \aj, 160, 193, \dodoi{10.3847/1538-3881/abb3bd}

\bibitem[{{Dawson}(2014)}]{dawson2014equil}
{Dawson}, R.~I. 2014, \apjl, 790, L31, \dodoi{10.1088/2041-8205/790/2/L31}

\bibitem[{{Dawson} \& {Johnson}(2018)}]{dawson2018hotjupiters}
{Dawson}, R.~I., \& {Johnson}, J.~A. 2018, \araa, 56, 175, \dodoi{10.1146/annurev-astro-081817-051853}

\bibitem[{{Do {\'O}} {et~al.}(2023){Do {\'O}}, {O'Neil}, {Konopacky}, {Do}, {Martinez}, {Ruffio}, \& {Ghez}}]{do2023priors}
{Do {\'O}}, C.~R., {O'Neil}, K.~K., {Konopacky}, Q.~M., {et~al.} 2023, \aj, 166, 48, \dodoi{10.3847/1538-3881/acdc9a}

\bibitem[{{Donati} {et~al.}(1997){Donati}, {Semel}, {Carter}, {Rees}, \& {Collier Cameron}}]{donati1997lsd}
{Donati}, J.~F., {Semel}, M., {Carter}, B.~D., {Rees}, D.~E., \& {Collier Cameron}, A. 1997, \mnras, 291, 658, \dodoi{10.1093/mnras/291.4.658}

\bibitem[{{Dorval} {et~al.}(2017){Dorval}, {Boily}, {Moraux}, \& {Roos}}]{dorval2017binary}
{Dorval}, J., {Boily}, C.~M., {Moraux}, E., \& {Roos}, O. 2017, \mnras, 465, 2198, \dodoi{10.1093/mnras/stw2880}

\bibitem[{{Durisen} {et~al.}(2007){Durisen}, {Boss}, {Mayer}, {Nelson}, {Quinn}, \& {Rice}}]{durisen2007formation}
{Durisen}, R.~H., {Boss}, A.~P., {Mayer}, L., {et~al.} 2007, in Protostars and Planets V, ed. B.~{Reipurth}, D.~{Jewitt}, \& K.~{Keil}, 607, \dodoi{10.48550/arXiv.astro-ph/0603179}

\bibitem[{{Fabrycky} \& {Tremaine}(2007)}]{fabrycky2007kozai}
{Fabrycky}, D., \& {Tremaine}, S. 2007, \apj, 669, 1298, \dodoi{10.1086/521702}

\bibitem[{{Ferreira} {et~al.}(2024){Ferreira}, {Rice}, {Wang}, \& {Wang}}]{ferreira2024toi2533}
{Ferreira}, T., {Rice}, M., {Wang}, X.-Y., \& {Wang}, S. 2024, arXiv e-prints, arXiv:2408.00725, \dodoi{10.48550/arXiv.2408.00725}

\bibitem[{{Fuller}(2017)}]{fuller2017resonancelock}
{Fuller}, J. 2017, \mnras, 472, 1538, \dodoi{10.1093/mnras/stx2135}

\bibitem[{{Gaia Collaboration} {et~al.}(2022){Gaia Collaboration}, {Vallenari}, {Brown}, {Prusti}, {de Bruijne}, {Arenou}, {Babusiaux}, {Biermann}, {Creevey}, {Ducourant}, {Evans}, {Eyer}, {Guerra}, {Hutton}, {Jordi}, {Klioner}, {Lammers}, {Lindegren}, {Luri}, {Mignard}, {Panem}, {Pourbaix}, {Randich}, {Sartoretti}, {Soubiran}, {Tanga}, {Walton}, {Bailer-Jones}, {Bastian}, {Drimmel}, {Jansen}, {Katz}, {Lattanzi}, {van Leeuwen}, {Bakker}, {Cacciari}, {Casta{\~n}eda}, {De Angeli}, {Fabricius}, {Fouesneau}, {Fr{\'e}mat}, {Galluccio}, {Guerrier}, {Heiter}, {Masana}, {Messineo}, {Mowlavi}, {Nicolas}, {Nienartowicz}, {Pailler}, {Panuzzo}, {Riclet}, {Roux}, {Seabroke}, {Sordo{\o}rcit}, {Th{\'e}venin}, {Gracia-Abril}, {Portell}, {Teyssier}, {Altmann}, {Andrae}, {Audard}, {Bellas-Velidis}, {Benson}, {Berthier}, {Blomme}, {Burgess}, {Busonero}, {Busso}, {C{\'a}novas}, {Carry}, {Cellino}, {Cheek}, {Clementini}, {Damerdji}, {Davidson}, {de Teodoro}, {Nu{\~n}ez Campos}, {Delchambre}, {Dell'Oro}, {Esquej},
  {Fern{\'a}ndez-Hern{\'a}ndez}, {Fraile}, {Garabato}, {Garc{\'\i}a-Lario}, {Gosset}, {Haigron}, {Halbwachs}, {Hambly}, {Harrison}, {Hern{\'a}ndez}, {Hestroffer}, {Hodgkin}, {Holl}, {Jan{\ss}en}, {Jevardat de Fombelle}, {Jordan}, {Krone-Martins}, {Lanzafame}, {L{\"o}ffler}, {Marchal}, {Marrese}, {Moitinho}, {Muinonen}, {Osborne}, {Pancino}, {Pauwels}, {Recio-Blanco}, {Reyl{\'e}}, {Riello}, {Rimoldini}, {Roegiers}, {Rybizki}, {Sarro}, {Siopis}, {Smith}, {Sozzetti}, {Utrilla}, {van Leeuwen}, {Abbas}, {{\'A}brah{\'a}m}, {Abreu Aramburu}, {Aerts}, {Aguado}, {Ajaj}, {Aldea-Montero}, {Altavilla}, {{\'A}lvarez}, {Alves}, {Anders}, {Anderson}, {Anglada Varela}, {Antoja}, {Baines}, {Baker}, {Balaguer-N{\'u}{\~n}ez}, {Balbinot}, {Balog}, {Barache}, {Barbato}, {Barros}, {Barstow}, {Bartolom{\'e}}, {Bassilana}, {Bauchet}, {Becciani}, {Bellazzini}, {Berihuete}, {Bernet}, {Bertone}, {Bianchi}, {Binnenfeld}, {Blanco-Cuaresma}, {Blazere}, {Boch}, {Bombrun}, {Bossini}, {Bouquillon}, {Bragaglia}, {Bramante}, {Breedt},
  {Bressan}, {Brouillet}, {Brugaletta}, {Bucciarelli}, {Burlacu}, {Butkevich}, {Buzzi}, {Caffau}, {Cancelliere}, {Cantat-Gaudin}, {Carballo}, {Carlucci}, {Carnerero}, {Carrasco}, {Casamiquela}, {Castellani}, {Castro-Ginard}, {Chaoul}, {Charlot}, {Chemin}, {Chiaramida}, {Chiavassa}, {Chornay}, {Comoretto}, {Contursi}, {Cooper}, {Cornez}, {Cowell}, {Crifo}, {Cropper}, {Crosta}, {Crowley}, {Dafonte}, {Dapergolas}, {David}, {David}, {de Laverny}, {De Luise}, {De March}, {De Ridder}, {de Souza}, {de Torres}, {del Peloso}, {del Pozo}, {Delbo}, {Delgado}, {Delisle}, {Demouchy}, {Dharmawardena}, {Di Matteo}, {Diakite}, {Diener}, {Distefano}, {Dolding}, {Edvardsson}, {Enke}, {Fabre}, {Fabrizio}, {Faigler}, {Fedorets}, {Fernique}, {Fienga}, {Figueras}, {Fournier}, {Fouron}, {Fragkoudi}, {Gai}, {Garcia-Gutierrez}, {Garcia-Reinaldos}, {Garc{\'\i}a-Torres}, {Garofalo}, {Gavel}, {Gavras}, {Gerlach}, {Geyer}, {Giacobbe}, {Gilmore}, {Girona}, {Giuffrida}, {Gomel}, {Gomez}, {Gonz{\'a}lez-N{\'u}{\~n}ez},
  {Gonz{\'a}lez-Santamar{\'\i}a}, {Gonz{\'a}lez-Vidal}, {Granvik}, {Guillout}, {Guiraud}, {Guti{\'e}rrez-S{\'a}nchez}, {Guy}, {Hatzidimitriou}, {Hauser}, {Haywood}, {Helmer}, {Helmi}, {Sarmiento}, {Hidalgo}, {Hilger}, {H{\l}adczuk}, {Hobbs}, {Holland}, {Huckle}, {Jardine}, {Jasniewicz}, {Jean-Antoine Piccolo}, {Jim{\'e}nez-Arranz}, {Jorissen}, {Juaristi Campillo}, {Julbe}, {Karbevska}, {Kervella}, {Khanna}, {Kontizas}, {Kordopatis}, {Korn}, {K{\'o}sp{\'a}l}, {Kostrzewa-Rutkowska}, {Kruszy{\'n}ska}, {Kun}, {Laizeau}, {Lambert}, {Lanza}, {Lasne}, {Le Campion}, {Lebreton}, {Lebzelter}, {Leccia}, {Leclerc}, {Lecoeur-Taibi}, {Liao}, {Licata}, {Lindstr{\o}m}, {Lister}, {Livanou}, {Lobel}, {Lorca}, {Loup}, {Madrero Pardo}, {Magdaleno Romeo}, {Managau}, {Mann}, {Manteiga}, {Marchant}, {Marconi}, {Marcos}, {Marcos Santos}, {Mar{\'\i}n Pina}, {Marinoni}, {Marocco}, {Marshall}, {Polo}, {Mart{\'\i}n-Fleitas}, {Marton}, {Mary}, {Masip}, {Massari}, {Mastrobuono-Battisti}, {Mazeh}, {McMillan}, {Messina}, {Michalik},
  {Millar}, {Mints}, {Molina}, {Molinaro}, {Moln{\'a}r}, {Monari}, {Mongui{\'o}}, {Montegriffo}, {Montero}, {Mor}, {Mora}, {Morbidelli}, {Morel}, {Morris}, {Muraveva}, {Murphy}, {Musella}, {Nagy}, {Noval}, {Oca{\~n}a}, {Ogden}, {Ordenovic}, {Osinde}, {Pagani}, {Pagano}, {Palaversa}, {Palicio}, {Pallas-Quintela}, {Panahi}, {Payne-Wardenaar}, {Pe{\~n}alosa Esteller}, {Penttil{\"a}}, {Pichon}, {Piersimoni}, {Pineau}, {Plachy}, {Plum}, {Poggio}, {Pr{\v{s}}a}, {Pulone}, {Racero}, {Ragaini}, {Rainer}, {Raiteri}, {Rambaux}, {Ramos}, {Ramos-Lerate}, {Re Fiorentin}, {Regibo}, {Richards}, {Rios Diaz}, {Ripepi}, {Riva}, {Rix}, {Rixon}, {Robichon}, {Robin}, {Robin}, {Roelens}, {Rogues}, {Rohrbasser}, {Romero-G{\'o}mez}, {Rowell}, {Royer}, {Ruz Mieres}, {Rybicki}, {Sadowski}, {S{\'a}ez N{\'u}{\~n}ez}, {Sagrist{\`a} Sell{\'e}s}, {Sahlmann}, {Salguero}, {Samaras}, {Sanchez Gimenez}, {Sanna}, {Santove{\~n}a}, {Sarasso}, {Schultheis}, {Sciacca}, {Segol}, {Segovia}, {S{\'e}gransan}, {Semeux}, {Shahaf}, {Siddiqui}, {Siebert},
  {Siltala}, {Silvelo}, {Slezak}, {Slezak}, {Smart}, {Snaith}, {Solano}, {Solitro}, {Souami}, {Souchay}, {Spagna}, {Spina}, {Spoto}, {Steele}, {Steidelm{\"u}ller}, {Stephenson}, {S{\"u}veges}, {Surdej}, {Szabados}, {Szegedi-Elek}, {Taris}, {Taylo}, {Teixeira}, {Tolomei}, {Tonello}, {Torra}, {Torra}, {Torralba Elipe}, {Trabucchi}, {Tsounis}, {Turon}, {Ulla}, {Unger}, {Vaillant}, {van Dillen}, {van Reeven}, {Vanel}, {Vecchiato}, {Viala}, {Vicente}, {Voutsinas}, {Weiler}, {Wevers}, {Wyrzykowski}, {Yoldas}, {Yvard}, {Zhao}, {Zorec}, {Zucker}, \& {Zwitter}}]{gaia2022}
{Gaia Collaboration}, {Vallenari}, A., {Brown}, A.~G.~A., {et~al.} 2022, arXiv e-prints, arXiv:2208.00211, \dodoi{10.48550/arXiv.2208.00211}

\bibitem[{{Gan} {et~al.}(2023){Gan}, {Wang}, {Wang}, {Mao}, {Huang}, {Collins}, {Stassun}, {Shporer}, {Zhu}, {Ricker}, {Vanderspek}, {Latham}, {Seager}, {Winn}, {Jenkins}, {Barkaoui}, {Belinski}, {Ciardi}, {Evans}, {Girardin}, {Maslennikova}, {Mazeh}, {Panahi}, {Pozuelos}, {Radford}, {Schwarz}, {Twicken}, {W{\"u}nsche}, \& {Zucker}}]{gan2023hotjupiter}
{Gan}, T., {Wang}, S.~X., {Wang}, S., {et~al.} 2023, \aj, 165, 17, \dodoi{10.3847/1538-3881/ac9b12}

\bibitem[{{Gibson} {et~al.}(2016){Gibson}, {Howard}, {Marcy}, {Edelstein}, {Wishnow}, \& {Poppett}}]{gibson2016kpf}
{Gibson}, S.~R., {Howard}, A.~W., {Marcy}, G.~W., {et~al.} 2016, in Society of Photo-Optical Instrumentation Engineers (SPIE) Conference Series, Vol. 9908, Ground-based and Airborne Instrumentation for Astronomy VI, ed. C.~J. {Evans}, L.~{Simard}, \& H.~{Takami}, 990870, \dodoi{10.1117/12.2233334}

\bibitem[{{Gibson} {et~al.}(2018){Gibson}, {Howard}, {Roy}, {Smith}, {Halverson}, {Edelstein}, {Kassis}, {Wishnow}, {Raffanti}, {Allen}, {Chin}, {Coutts}, {Cowley}, {Curtis}, {Deich}, {Feger}, {Finstad}, {Gurevich}, {Ishikawa}, {James}, {Jhoti}, {Lanclos}, {Lilley}, {Miller}, {Milner}, {Payne}, {Rider}, {Rockosi}, {Sandford}, {Schwab}, {Seifahrt}, {Sirk}, {Smith}, {Stuermer}, {Weisfeiler}, {Wilcox}, {Vandenberg}, \& {Wizinowich}}]{gibson2018kpf}
{Gibson}, S.~R., {Howard}, A.~W., {Roy}, A., {et~al.} 2018, in Society of Photo-Optical Instrumentation Engineers (SPIE) Conference Series, Vol. 10702, Ground-based and Airborne Instrumentation for Astronomy VII, ed. C.~J. {Evans}, L.~{Simard}, \& H.~{Takami}, 107025X, \dodoi{10.1117/12.2311565}

\bibitem[{{Gibson} {et~al.}(2020){Gibson}, {Howard}, {Rider}, {Roy}, {Edelstein}, {Kassis}, {Grillo}, {Halverson}, {Sirk}, {Smith}, {Allen}, {Baker}, {Beichman}, {Berriman}, {Brown}, {Casey}, {Chin}, {Coutts}, {Cowley}, {Deich}, {Feger}, {Fulton}, {Gers}, {Gurevich}, {Ishikawa}, {James}, {Jelinsky}, {Kaye}, {Lanclos}, {Li}, {Lilley}, {McCarney}, {Miller}, {Milner}, {O'Hanlon}, {Pember}, {Raffanti}, {Rockosi}, {Rubenzahl}, {Rumph}, {Sandford}, {Savage}, {Schwab}, {Seifahrt}, {Shaum}, {Smith}, {Stuermer}, {Thorne}, {Vandenberg}, {Von Boeckmann}, {Wang}, {Wang}, {Weisfeiler}, {Wilcox}, {Wishnow}, {Wizinowich}, {Wold}, \& {Wolfenberger}}]{gibson2020kpf}
{Gibson}, S.~R., {Howard}, A.~W., {Rider}, K., {et~al.} 2020, in Society of Photo-Optical Instrumentation Engineers (SPIE) Conference Series, Vol. 11447, Ground-based and Airborne Instrumentation for Astronomy VIII, ed. C.~J. {Evans}, J.~J. {Bryant}, \& K.~{Motohara}, 1144742, \dodoi{10.1117/12.2561783}

\bibitem[{{Grether} \& {Lineweaver}(2006)}]{grether2006desert}
{Grether}, D., \& {Lineweaver}, C.~H. 2006, \apj, 640, 1051, \dodoi{10.1086/500161}

\bibitem[{{Grieves} {et~al.}(2021){Grieves}, {Bouchy}, {Lendl}, {Carmichael}, {Mireles}, {Shporer}, {McLeod}, {Collins}, {Brahm}, {Stassun}, {Gill}, {Bouma}, {Guillot}, {Cointepas}, {Dos Santos}, {Casewell}, {Jenkins}, {Henning}, {Nielsen}, {Psaridi}, {Udry}, {S{\'e}gransan}, {Eastman}, {Zhou}, {Abe}, {Agabi}, {Bakos}, {Charbonneau}, {Collins}, {Colon}, {Crouzet}, {Dransfield}, {Evans}, {Goeke}, {Hart}, {Irwin}, {Jensen}, {Jord{\'a}n}, {Kielkopf}, {Latham}, {Marie-Sainte}, {M{\'e}karnia}, {Nelson}, {Quinn}, {Radford}, {Rodriguez}, {Rowden}, {Schmider}, {Schwarz}, {Smith}, {Stockdale}, {Suarez}, {Tan}, {Triaud}, {Waalkes}, \& {Wingham}}]{grieves2021bd}
{Grieves}, N., {Bouchy}, F., {Lendl}, M., {et~al.} 2021, \aap, 652, A127, \dodoi{10.1051/0004-6361/202141145}

\bibitem[{{Howard} {et~al.}(2012){Howard}, {Marcy}, {Bryson}, {Jenkins}, {Rowe}, {Batalha}, {Borucki}, {Koch}, {Dunham}, {Gautier}, {Van Cleve}, {Cochran}, {Latham}, {Lissauer}, {Torres}, {Brown}, {Gilliland}, {Buchhave}, {Caldwell}, {Christensen-Dalsgaard}, {Ciardi}, {Fressin}, {Haas}, {Howell}, {Kjeldsen}, {Seager}, {Rogers}, {Sasselov}, {Steffen}, {Basri}, {Charbonneau}, {Christiansen}, {Clarke}, {Dupree}, {Fabrycky}, {Fischer}, {Ford}, {Fortney}, {Tarter}, {Girouard}, {Holman}, {Johnson}, {Klaus}, {Machalek}, {Moorhead}, {Morehead}, {Ragozzine}, {Tenenbaum}, {Twicken}, {Quinn}, {Isaacson}, {Shporer}, {Lucas}, {Walkowicz}, {Welsh}, {Boss}, {Devore}, {Gould}, {Smith}, {Morris}, {Prsa}, {Morton}, {Still}, {Thompson}, {Mullally}, {Endl}, \& {MacQueen}}]{howard2012kepler}
{Howard}, A.~W., {Marcy}, G.~W., {Bryson}, S.~T., {et~al.} 2012, \apjs, 201, 15, \dodoi{10.1088/0067-0049/201/2/15}

\bibitem[{{Howell} {et~al.}(2014){Howell}, {Sobeck}, {Haas}, {Still}, {Barclay}, {Mullally}, {Troeltzsch}, {Aigrain}, {Bryson}, {Caldwell}, {Chaplin}, {Cochran}, {Huber}, {Marcy}, {Miglio}, {Najita}, {Smith}, {Twicken}, \& {Fortney}}]{howell2014k2}
{Howell}, S.~B., {Sobeck}, C., {Haas}, M., {et~al.} 2014, \pasp, 126, 398, \dodoi{10.1086/676406}

\bibitem[{{Hurley} {et~al.}(2002){Hurley}, {Tout}, \& {Pols}}]{Hurley+(2002)}
{Hurley}, J.~R., {Tout}, C.~A., \& {Pols}, O.~R. 2002, \mnras, 329, 897, \dodoi{10.1046/j.1365-8711.2002.05038.x}

\bibitem[{{Husnoo} {et~al.}(2012){Husnoo}, {Pont}, {Mazeh}, {Fabrycky}, {H{\'e}brard}, {Bouchy}, \& {Shporer}}]{husnoo2012tides}
{Husnoo}, N., {Pont}, F., {Mazeh}, T., {et~al.} 2012, \mnras, 422, 3151, \dodoi{10.1111/j.1365-2966.2012.20839.x}

\bibitem[{{Hut}(1980)}]{hut1980equil}
{Hut}, P. 1980, \aap, 92, 167

\bibitem[{{Kushnir} {et~al.}(2017){Kushnir}, {Zaldarriaga}, {Kollmeier}, \& {Waldman}}]{Kushnir+(2017)}
{Kushnir}, D., {Zaldarriaga}, M., {Kollmeier}, J.~A., \& {Waldman}, R. 2017, \mnras, 467, 2146, \dodoi{10.1093/mnras/stx255}

\bibitem[{{Lin} {et~al.}(2023){Lin}, {Gan}, {Wang}, {Shporer}, {Rabus}, {Zhou}, {Psaridi}, {Bouchy}, {Bieryla}, {Latham}, {Mao}, {Stassun}, {Hellier}, {Howell}, {Ziegler}, {Caldwell}, {Clark}, {Collins}, {Curtis}, {Faherty}, {Gnilka}, {Grunblatt}, {Jenkins}, {Johnson}, {Law}, {Lendl}, {Littlefield}, {Lund}, {Lund}, {Mann}, {McDermott}, {Mishra}, {Mounzer}, {Paegert}, {Pritchard}, {Ricker}, {Seager}, {Srdoc}, {Sun}, {Tang}, {Udry}, {Vanderspek}, {Watanabe}, {Winn}, \& {Yu}}]{lin2023bd}
{Lin}, Z., {Gan}, T., {Wang}, S.~X., {et~al.} 2023, \mnras, 523, 6162, \dodoi{10.1093/mnras/stad1745}

\bibitem[{{Ma} \& {Ge}(2014)}]{ma2014statistical}
{Ma}, B., \& {Ge}, J. 2014, \mnras, 439, 2781, \dodoi{10.1093/mnras/stu134}

\bibitem[{{Ma} \& {Fuller}(2021)}]{ma2021resonancelock}
{Ma}, L., \& {Fuller}, J. 2021, \apj, 918, 16, \dodoi{10.3847/1538-4357/ac088e}

\bibitem[{{Matsumura} {et~al.}(2010){Matsumura}, {Peale}, \& {Rasio}}]{matsumura2010tidal}
{Matsumura}, S., {Peale}, S.~J., \& {Rasio}, F.~A. 2010, \apj, 725, 1995, \dodoi{10.1088/0004-637X/725/2/1995}

\bibitem[{{Matsuo} {et~al.}(2007){Matsuo}, {Shibai}, {Ootsubo}, \& {Tamura}}]{matsuo2007formation}
{Matsuo}, T., {Shibai}, H., {Ootsubo}, T., \& {Tamura}, M. 2007, \apj, 662, 1282, \dodoi{10.1086/517964}

\bibitem[{{McLaughlin}(1924)}]{mclaughlin1924rm}
{McLaughlin}, D.~B. 1924, \apj, 60, 22, \dodoi{10.1086/142826}

\bibitem[{{Nagpal} {et~al.}(2023){Nagpal}, {Blunt}, {Bowler}, {Dupuy}, {Nielsen}, \& {Wang}}]{nagpal2023priors}
{Nagpal}, V., {Blunt}, S., {Bowler}, B.~P., {et~al.} 2023, \aj, 165, 32, \dodoi{10.3847/1538-3881/ac9fd2}

\bibitem[{{Nielsen} {et~al.}(2019){Nielsen}, {De Rosa}, {Macintosh}, {Wang}, {Ruffio}, {Chiang}, {Marley}, {Saumon}, {Savransky}, {Ammons}, {Bailey}, {Barman}, {Blain}, {Bulger}, {Burrows}, {Chilcote}, {Cotten}, {Czekala}, {Doyon}, {Duch{\^e}ne}, {Esposito}, {Fabrycky}, {Fitzgerald}, {Follette}, {Fortney}, {Gerard}, {Goodsell}, {Graham}, {Greenbaum}, {Hibon}, {Hinkley}, {Hirsch}, {Hom}, {Hung}, {Dawson}, {Ingraham}, {Kalas}, {Konopacky}, {Larkin}, {Lee}, {Lin}, {Maire}, {Marchis}, {Marois}, {Metchev}, {Millar-Blanchaer}, {Morzinski}, {Oppenheimer}, {Palmer}, {Patience}, {Perrin}, {Poyneer}, {Pueyo}, {Rafikov}, {Rajan}, {Rameau}, {Rantakyr{\"o}}, {Ren}, {Schneider}, {Sivaramakrishnan}, {Song}, {Soummer}, {Tallis}, {Thomas}, {Ward-Duong}, \& {Wolff}}]{nielsen2019gemini}
{Nielsen}, E.~L., {De Rosa}, R.~J., {Macintosh}, B., {et~al.} 2019, \aj, 158, 13, \dodoi{10.3847/1538-3881/ab16e9}

\bibitem[{{Offner} {et~al.}(2016){Offner}, {Dunham}, {Lee}, {Arce}, \& {Fielding}}]{offner2016misalignment}
{Offner}, S. S.~R., {Dunham}, M.~M., {Lee}, K.~I., {Arce}, H.~G., \& {Fielding}, D.~B. 2016, \apjl, 827, L11, \dodoi{10.3847/2041-8205/827/1/L11}

\bibitem[{{Petrovich}(2015)}]{petrovich2015coplanar}
{Petrovich}, C. 2015, \apj, 805, 75, \dodoi{10.1088/0004-637X/805/1/75}

\bibitem[{{Pollack} {et~al.}(1996){Pollack}, {Hubickyj}, {Bodenheimer}, {Lissauer}, {Podolak}, \& {Greenzweig}}]{pollack1996coreaccretion}
{Pollack}, J.~B., {Hubickyj}, O., {Bodenheimer}, P., {et~al.} 1996, \icarus, 124, 62, \dodoi{10.1006/icar.1996.0190}

\bibitem[{{Psaridi} {et~al.}(2022){Psaridi}, {Bouchy}, {Lendl}, {Grieves}, {Stassun}, {Carmichael}, {Gill}, {Pe{\~n}a Rojas}, {Gan}, {Shporer}, {Bieryla}, {Brahm}, {Christiansen}, {Crossfield}, {Galland}, {Hooton}, {Jenkins}, {Jenkins}, {Latham}, {Lund}, {Rodriguez}, {Ting}, {Udry}, {Ulmer-Moll}, {Wittenmyer}, {Zhang}, {Zhou}, {Addison}, {Cointepas}, {Collins}, {Collins}, {Deline}, {Dressing}, {Evans}, {Giacalone}, {Heitzmann}, {Mireles}, {Mounzer}, {Otegi}, {Radford}, {Rudat}, {Schlieder}, {Schwarz}, {Srdoc}, {Stockdale}, {Suarez}, {Wright}, \& {Zhao}}]{psaridi2022bd}
{Psaridi}, A., {Bouchy}, F., {Lendl}, M., {et~al.} 2022, \aap, 664, A94, \dodoi{10.1051/0004-6361/202243454}

\bibitem[{{Rice} {et~al.}(2022){Rice}, {Wang}, \& {Laughlin}}]{rice2022obliquities}
{Rice}, M., {Wang}, S., \& {Laughlin}, G. 2022, \apjl, 926, L17, \dodoi{10.3847/2041-8213/ac502d}

\bibitem[{{Ricker} {et~al.}(2015){Ricker}, {Winn}, {Vanderspek}, {Latham}, {Bakos}, {Bean}, {Berta-Thompson}, {Brown}, {Buchhave}, {Butler}, {Butler}, {Chaplin}, {Charbonneau}, {Christensen-Dalsgaard}, {Clampin}, {Deming}, {Doty}, {De Lee}, {Dressing}, {Dunham}, {Endl}, {Fressin}, {Ge}, {Henning}, {Holman}, {Howard}, {Ida}, {Jenkins}, {Jernigan}, {Johnson}, {Kaltenegger}, {Kawai}, {Kjeldsen}, {Laughlin}, {Levine}, {Lin}, {Lissauer}, {MacQueen}, {Marcy}, {McCullough}, {Morton}, {Narita}, {Paegert}, {Palle}, {Pepe}, {Pepper}, {Quirrenbach}, {Rinehart}, {Sasselov}, {Sato}, {Seager}, {Sozzetti}, {Stassun}, {Sullivan}, {Szentgyorgyi}, {Torres}, {Udry}, \& {Villasenor}}]{ricker2015tess}
{Ricker}, G.~R., {Winn}, J.~N., {Vanderspek}, R., {et~al.} 2015, Journal of Astronomical Telescopes, Instruments, and Systems, 1, 014003, \dodoi{10.1117/1.JATIS.1.1.014003}

\bibitem[{{Rossiter}(1924)}]{rossiter1924rm}
{Rossiter}, R.~A. 1924, \apj, 60, 15, \dodoi{10.1086/142825}

\bibitem[{{Ryabchikova} {et~al.}(2015){Ryabchikova}, {Piskunov}, {Kurucz}, {Stempels}, {Heiter}, {Pakhomov}, \& {Barklem}}]{ryabchikova2015vald3}
{Ryabchikova}, T., {Piskunov}, N., {Kurucz}, R.~L., {et~al.} 2015, \physscr, 90, 054005, \dodoi{10.1088/0031-8949/90/5/054005}

\bibitem[{{Savonije}(2008)}]{savonije2008resonancelock}
{Savonije}, G.~J. 2008, in EAS Publications Series, Vol.~29, EAS Publications Series, ed. M.~J. {Goupil} \& J.~P. {Zahn}, 91--125, \dodoi{10.1051/eas:0829003}

\bibitem[{{Schlaufman}(2018)}]{schlaufman2018evidence}
{Schlaufman}, K.~C. 2018, \apj, 853, 37, \dodoi{10.3847/1538-4357/aa961c}

\bibitem[{{Siverd} {et~al.}(2012){Siverd}, {Beatty}, {Pepper}, {Eastman}, {Collins}, {Bieryla}, {Latham}, {Buchhave}, {Jensen}, {Crepp}, {Street}, {Stassun}, {Gaudi}, {Berlind}, {Calkins}, {DePoy}, {Esquerdo}, {Fulton}, {F{\H{u}}r{\'e}sz}, {Geary}, {Gould}, {Hebb}, {Kielkopf}, {Marshall}, {Pogge}, {Stanek}, {Stefanik}, {Szentgyorgyi}, {Trueblood}, {Trueblood}, {Stutz}, \& {van Saders}}]{siverd2012kelt1}
{Siverd}, R.~J., {Beatty}, T.~G., {Pepper}, J., {et~al.} 2012, \apj, 761, 123, \dodoi{10.1088/0004-637X/761/2/123}

\bibitem[{{Southworth}(2011)}]{southworth2011tepcat}
{Southworth}, J. 2011, \mnras, 417, 2166, \dodoi{10.1111/j.1365-2966.2011.19399.x}

\bibitem[{{Speagle}(2020)}]{Speagle}
{Speagle}, J.~S. 2020, \mnras, 493, 3132, \dodoi{10.1093/mnras/staa278}

\bibitem[{{Spiegel} {et~al.}(2011){Spiegel}, {Burrows}, \& {Milsom}}]{spiegel2011deuterium}
{Spiegel}, D.~S., {Burrows}, A., \& {Milsom}, J.~A. 2011, \apj, 727, 57, \dodoi{10.1088/0004-637X/727/1/57}

\bibitem[{{Su} \& {Lai}(2022)}]{SuLai(2022)}
{Su}, Y., \& {Lai}, D. 2022, \mnras, 510, 4943, \dodoi{10.1093/mnras/stab3698}

\bibitem[{{Tokovinin} \& {Moe}(2020)}]{tokovinin2020migration}
{Tokovinin}, A., \& {Moe}, M. 2020, \mnras, 491, 5158, \dodoi{10.1093/mnras/stz3299}

\bibitem[{{Triaud} {et~al.}(2009){Triaud}, {Queloz}, {Bouchy}, {Moutou}, {Collier Cameron}, {Claret}, {Barge}, {Benz}, {Deleuil}, {Guillot}, {H{\'e}brard}, {Lecavelier Des {\'E}tangs}, {Lovis}, {Mayor}, {Pepe}, \& {Udry}}]{triaud2009corot3}
{Triaud}, A.~H.~M.~J., {Queloz}, D., {Bouchy}, F., {et~al.} 2009, \aap, 506, 377, \dodoi{10.1051/0004-6361/200911897}

\bibitem[{{Triaud} {et~al.}(2013){Triaud}, {Hebb}, {Anderson}, {Cargile}, {Collier Cameron}, {Doyle}, {Faedi}, {Gillon}, {Gomez Maqueo Chew}, {Hellier}, {Jehin}, {Maxted}, {Naef}, {Pepe}, {Pollacco}, {Queloz}, {S{\'e}gransan}, {Smalley}, {Stassun}, {Udry}, \& {West}}]{triaud2013wasp30}
{Triaud}, A.~H.~M.~J., {Hebb}, L., {Anderson}, D.~R., {et~al.} 2013, \aap, 549, A18, \dodoi{10.1051/0004-6361/201219643}

\bibitem[{{Vowell} {et~al.}(2023){Vowell}, {Rodriguez}, {Quinn}, {Zhou}, {Vanderburg}, {Mann}, {Hooton}, {Stassun}, {Howard}, {Bieryla}, {Latham}, {Howell}, {Guillot}, {Ziegler}, {Collins}, {Carmichael}, {Jenkins}, {Shporer}, {ABE}, {Bendjoya}, {Bush}, {Buttu}, {Collins}, {Eastman}, {Fields}, {Gasparetto}, {G{\"u}nther}, {Kostov}, {Kraus}, {Lester}, {Levine}, {Littlefield}, {Marie-Sainte}, {M{\'e}karnia}, {Osborn}, {Rapetti}, {Ricker}, {Seager}, {Sefako}, {Srdoc}, {Suarez}, {Torres}, {Triaud}, {Vanderspek}, \& {Winn}}]{vowell2023bd}
{Vowell}, N., {Rodriguez}, J.~E., {Quinn}, S.~N., {et~al.} 2023, \aj, 165, 268, \dodoi{10.3847/1538-3881/acd197}

\bibitem[{{Winn} {et~al.}(2010){Winn}, {Fabrycky}, {Albrecht}, \& {Johnson}}]{winn2010obliquity}
{Winn}, J.~N., {Fabrycky}, D., {Albrecht}, S., \& {Johnson}, J.~A. 2010, \apjl, 718, L145, \dodoi{10.1088/2041-8205/718/2/L145}

\bibitem[{{Zahn}(1975)}]{zahn1975dynamicaltide}
{Zahn}, J.~P. 1975, \aap, 41, 329

\bibitem[{{Zahn}(1977)}]{zahn1977tidal}
---. 1977, \aap, 57, 383

\bibitem[{{Zanazzi} {et~al.}(2024){Zanazzi}, {Dewberry}, \& {Chiang}}]{zanazzi2024damping}
{Zanazzi}, J.~J., {Dewberry}, J., \& {Chiang}, E. 2024, \apjl, 967, L29, \dodoi{10.3847/2041-8213/ad4644}

\bibitem[{{Zanazzi} \& {Wu}(2021)}]{zanazzi2021resonancelock}
{Zanazzi}, J.~J., \& {Wu}, Y. 2021, \aj, 161, 263, \dodoi{10.3847/1538-3881/abf097}

\bibitem[{{Zhou} {et~al.}(2019{\natexlab{a}}){Zhou}, {Huang}, {Bakos}, {Hartman}, {Latham}, {Quinn}, {Collins}, {Winn}, {Wong}, {Kov{\'a}cs}, {Csubry}, {Bhatti}, {Penev}, {Bieryla}, {Esquerdo}, {Berlind}, {Calkins}, {de Val-Borro}, {Noyes}, {L{\'a}z{\'a}r}, {Papp}, {S{\'a}ri}, {Kov{\'a}cs}, {Buchhave}, {Szklenar}, {B{\'e}ky}, {Johnson}, {Cochran}, {Kniazev}, {Stassun}, {Fulton}, {Shporer}, {Espinoza}, {Bayliss}, {Everett}, {Howell}, {Hellier}, {Anderson}, {Collier Cameron}, {West}, {Brown}, {Schanche}, {Barkaoui}, {Pozuelos}, {Gillon}, {Jehin}, {Benkhaldoun}, {Daassou}, {Ricker}, {Vanderspek}, {Seager}, {Jenkins}, {Lissauer}, {Armstrong}, {Collins}, {Gan}, {Hart}, {Horne}, {Kielkopf}, {Nielsen}, {Nishiumi}, {Narita}, {Palle}, {Relles}, {Sefako}, {Tan}, {Davies}, {Goeke}, {Guerrero}, {Haworth}, \& {Villanueva}}]{zhou2019hotjupiter}
{Zhou}, G., {Huang}, C.~X., {Bakos}, G.~{\'A}., {et~al.} 2019{\natexlab{a}}, \aj, 158, 141, \dodoi{10.3847/1538-3881/ab36b5}

\bibitem[{{Zhou} {et~al.}(2019{\natexlab{b}}){Zhou}, {Bakos}, {Bayliss}, {Bento}, {Bhatti}, {Brahm}, {Csubry}, {Espinoza}, {Hartman}, {Henning}, {Jord{\'a}n}, {Mancini}, {Penev}, {Rabus}, {Sarkis}, {Suc}, {de Val-Borro}, {Rodriguez}, {Osip}, {Kedziora-Chudczer}, {Bailey}, {Tinney}, {Durkan}, {L{\'a}z{\'a}r}, {Papp}, \& {S{\'a}ri}}]{zhou2019hats70}
{Zhou}, G., {Bakos}, G.~{\'A}., {Bayliss}, D., {et~al.} 2019{\natexlab{b}}, \aj, 157, 31, \dodoi{10.3847/1538-3881/aaf1bb}

\end{thebibliography}
\bibliographystyle{aasjournal}

\end{document}